\documentclass[usenatbib,onecolumn]{mn2e}
\usepackage{natbib}
\usepackage{epsfig}
\usepackage{lscape}
\usepackage{psfig}

\catcode`\@=11
\def\gta{\ifmmode{\,\mathrel{\mathpalette\@versim>\,}}
    \else{$\,\mathrel{\mathpalette\@versim>}\,$}\fi}
\def\lta{\ifmmode{\,\mathrel{\mathpalette\@versim<\,}}
    \else{$\,\mathrel{\mathpalette\@versim<}\,$}\fi}
\def\@versim#1#2{\lower 2.9truept \vbox{\baselineskip 0pt \lineskip
    0.5truept \ialign{$\m@th#1\hfil##\hfil$\crcr#2\crcr\sim\crcr}}}
\catcode`\@=12  

\def\ArhoR{A_{\rho{R}}}
\def\Arhoz{A_{\rho{z}}}
\def\ApR{A_{pR}}
\def\Apz{A_{pz}}
\def\czero{c_0}
\def\czerotil{\tilde{c}_0}
\def\d{{\rm d}}
\def\Dm{\mathcal{D}}
\def\Fzero{{F_0}}
\def\grad{{\nabla}}
\def\hi{H\,{\sc i}}
\def\i{{\rm i}}

\def\kz{k_z}
\def\kR{k_R}
\def\kztil{\tilde{k}_z}
\def\kRtil{\tilde{k}_R}
\def\ktil{\tilde{k}}
\def\kphi{k_{\phi}}
\def\kv{{\bf k}}
\def\kzksq{{\kz^2\over k^2}}
\def\mp{m_{\rm p}}

\def\ntilde{\tilde{n}}
\def\pd{\partial}
\def\pzero{{p_0}}
\def\szero{{s_0}}
\def\Tzero{T_0}
\def\vz{{v_z}}
\def\vR{{v_R}}
\def\vphi{v_{\phi}}
\def\vRtil{{\tilde{v}_R}}
\def\vztil{{\tilde{v}_z}}

\def\vvzero{\vv_0}
\def\vzerophi{v_{0\phi}}
\def\vzeroR{v_{0R}}
\def\vzeroz{v_{0z}}

\def\vv{{\bf v}}
\def\gammaprime{{\gamma^{\prime}}}
\def\GammapR{\Gamma_{pR}}
\def\Gammapz{\Gamma_{pz}}
\def\GammarhoR{\Gamma_{\rho{R}}}
\def\Gammarhoz{\Gamma_{\rho{z}}}
\def\GammaOmegaR{\Gamma_{\Omega{R}}}
\def\GammaOmegaz{\Gamma_{\Omega{z}}}
\def\rhozero{\rho_0}
\def\rhotil{\tilde{\rho}}
\def\omegad{\omega_{\rm d}}
\def\omegadtil{\tilde{\omega}_{\rm d}}
\def\omegadsq{\omega_{\rm d}^2}
\def\omegaBV{\omega_{\rm BV}}
\def\omegaBVsq{\omega^2_{\rm BV}}
\def\omegaBVsqtil{\tilde{\omega}^2_{\rm BV}}
\def\omegaBVtil{\tilde{\omega}_{\rm BV}}
\def\omegarot{\omega_{\rm rot}}
\def\omegarotsq{\omega^2_{\rm rot}}
\def\omegarotsqtil{\tilde{\omega}^2_{\rm rot}}
\def\omegarottil{\tilde{\omega}_{\rm rot}}
\def\omegac{\omega_{\rm c}}
\def\omegactil{\tilde{\omega}_{\rm c}}
\def\omegahat{\hat{\omega}}
\def\omegabar{\bar{\omega}}
\def\Omegaz{\Omega_{z}}
\def\OmegaR{\Omega_{R}}
\def\omegath{\omega_{\rm th}}
\def\omegathtil{\tilde{\omega}_{\rm th}}

\begin{document}
\date{Accepted 2010 March 15. Received 2010 March 10; in original form 2010 January 21}
\title{Thermal instability in rotating galactic coronae} 
\author[C. Nipoti]{Carlo Nipoti
\\
Dipartimento di Astronomia, Universit\`a di Bologna, via Ranzani 1, I-40127 Bologna, Italy}

\maketitle

\begin{abstract}
  The thermal stability of rotating, stratified, unmagnetized
  atmospheres is studied by means of linear-perturbation analysis,
  finding stability, overstability or instability, depending on the
  properties of the gas distribution, but also on the nature of the
  perturbations.  In the relevant case of distributions with
  outward-increasing specific entropy and angular momentum,
  axisymmetric perturbations grow exponentially, unless their
  wavelength is short enough that they are damped by thermal
  conduction; non-axisymmetric perturbations typically undergo
  overstable oscillations in the limit of zero conductivity, but are
  effectively stabilized by thermal conduction, provided rotation is
  differential. To the extent that the studied models are
  representative of the poorly constrained hot atmospheres of disc
  galaxies, these results imply that blob-like, cool overdensities are
  unlikely to grow in galactic coronae, suggesting an external origin
  for the high-velocity clouds of the Milky Way.
\end{abstract}


\section{Introduction}

Galaxy clusters and massive elliptical galaxies are embedded in hot
atmospheres of virial-temperature gas revealed by X-ray
observations. Lower-mass galaxies are also believed to have hot
gaseous coronae, difficult to detect because the gas is very
rarefied. If the coronal gas were thermally unstable \citep{Fie65} it
could fragment into cold gas clouds with substantial implications for
the dynamics of cooling flows \citep{Mat78,Cow80,Nul86}, but also for
galaxy formation and for the origin of the \hi\ high-velocity clouds
of the Milky Way \citep{Mal04,Pee08,Kau09}. In fact,
linear-perturbation analysis has shown that the X-ray emitting hot
atmospheres of massive elliptical galaxies and galaxy clusters are
likely stabilized against thermal instability by a combination of
buoyancy and thermal conduction \citep[Malagoli, Rosner \& Bodo 1987,
hereafter MRB;][]{BS89,Tribble89}.  The coronae of disc galaxies are
expected to have lower gas temperatures and densities, but their
detailed properties are poorly known, because---so far---they eluded
detection in X-rays. Nevertheless, their physical parameters are not
totally unconstrained: for some nearby massive disc galaxies upper
limits to the total X-ray luminosity of the hot halos are available
\citep{Ras09}, while in the special case of the Milky Way different
pieces of information can be used to constrain the physical properties
of the corona \citep[e.g.][]{Spi56,Fuk06}. Studying model coronae
consistent with these constraints, \citet[][hereafter BNF]{Bin09}
showed that also the coronae of disc galaxies are likely thermally
stabilized by buoyancy and thermal conduction, at least if rotation of
the gas can be neglected.  There are no significant constraints on the
angular momentum of galactic coronae, but one might expect that they
are characterized by at least slow rotation. It is well known that
there is an important interplay between rotation and convection
\citep[e.g.][]{Tas78,Bal00,Bal01}, so rotation might influence the
stabilizing effect of buoyancy and then the overall problem of the
thermal stability of galactic coronae.

This paper addresses the question of the thermal stability of rotating
galactic coronae. For this purpose one needs to study the stability
against non-axisymmetric perturbations of a differentially rotating,
stratified gas in the presence of thermal conduction and radiative
cooling.  This is a special case of the general problem of the
stability of a rotating stratified gas, which has been widely studied
in the astrophysical literature, with specific applications to
rotating
stars~\citep[e.g.][]{Cow51,Gol67,Leb67,Fri68,Lyn67,Tas78,Kno82,Lif93,Men04}
and accretion
discs~\citep[e.g.][]{Pap84,BalH91,BalH92,Ryu92,Lin93,Bra06}.  The
stability-analysis techniques developed in these previous studies can
be applied to the problem addressed in the present paper.  In
particular, we will perform a linear analysis using Eulerian
perturbations, considering both axisymmetric and non-axisymmetric
disturbances.

It is reasonable to expect that galactic coronae are magnetized,
though there are almost no constraints on the properties of their
magnetic fields. For simplicity, in the present work we neglect
magnetic fields, but we must bear in mind that they might have some
influence on the thermal stability properties of the galactic hot gas,
as suggested by previous studies of thermal instability in
non-rotating cooling flows \citep[][see also BNF for a
discussion]{Loe90,Bal91}.


The paper is organized as follows. In Section~\ref{sec:eq} the problem
is set up by recalling the relevant hydrodynamic equations.  The
thermal-stability analysis is then described for axisymmetric
perturbations (Section~\ref{sec:axisymm}) and for non-axisymmetric
perturbations (Section~\ref{sec:nonaxisymm}). The implications for
galactic coronae are discussed in Section~\ref{sec:impl};
Section~\ref{sec:con} concludes. A list of commonly used symbols is
given for reference in Table~\ref{tab:list}.

\section{Thermal instability in a rotating stratified gas: governing
  equations}

\label{sec:eq}

A rotating, stratified, unmagnetized atmosphere in the presence of
cooling and thermal conduction is governed by the following equations
for mass, momentum and energy conservation:
\begin{eqnarray}
&&{\partial \rho \over \partial t}+\grad\cdot(\rho\vv)=0,\\
&&{\partial \vv \over \partial t}+\vv\cdot\grad\vv=-{\grad p \over \rho}-\grad\Phi,\\
&&{p\over \gamma-1}\left[{\partial \over \partial
  t}+\vv\cdot\grad\right] \ln (p \rho^{-\gamma})=
\grad\cdot\left(\kappa T^{5/2}{\grad} T\right)-\left({\rho\over \mu \mp}\right)^2\Lambda(T). \label{eq:hydro}
\end{eqnarray}
Here $\rho$, $p$, $T$ and $\vv$ are, respectively, the gas density,
pressure, temperature and velocity, $\Phi$ is the galactic
gravitational potential, $\gamma=5/3$ is the ratio of principal
specific heats, $\Lambda$ is the cooling function, $\mu$ is the mean
gas particle mass in units of the proton mass $\mp$ and $\kappa$ is
the thermal conductivity \citep{Spi62}.  As we neglect magnetic
fields, we treat the thermal conductivity as a scalar. In this
approximation we can at most assume that the actual value of $\kappa$
is suppressed by tangled magnetic fields below Spitzer's benchmark
value \citep[e.g.][]{Bin81}, but of course we cannot account for the
effects related to the anisotropic heat transport expected in a
magnetized medium \citep[e.g.][]{Bal00,Qua08}.

  In cylindrical coordinates, assuming axisymmetric gravitational
  potential, the above hydrodynamic equations write:
\begin{eqnarray} 
&&{\partial \rho \over \partial t}+{1\over R}{\partial
    R\rho \vR \over \partial R} +{\partial \rho \vz \over \partial
    z}+{1\over R}{\partial \rho \vphi \over \partial \phi}=0,\label{eq:hydrocylc}\\
&&{\partial \vR \over \partial t}+\vR{\partial \vR \over \partial
    R}+\vz{\partial \vR \over \partial z}+{\vphi\over R}{\partial \vR
    \over \partial \phi}=-{1\over \rho}{\partial p \over \partial
    R}-{\partial \Phi \over \partial R}+{\vphi^2\over R},\\
&&{\partial \vz \over \partial t}+\vR{\partial \vz \over \partial
    R}+\vz{\partial \vz \over \partial z}+{\vphi\over R}{\partial \vz
    \over \partial \phi}=-{1\over \rho}{\partial p \over \partial
    z}-{\partial \Phi \over \partial z},\\ 
&&{\partial \vphi \over \partial t}+\vR{\partial \vphi \over \partial
    R}+\vz{\partial \vphi \over \partial z}+{\vphi\over R}{\partial
    \vphi \over \partial \phi}=-{1\over \rho R}{\partial p
    \over \partial \phi}-{\vR\vphi\over R},\\
&&{p\over \gamma-1}\left[{\partial \over \partial t}+\vR{\partial \over \partial  R}+\vz{\partial \over \partial  z}+{\vphi \over R}{\partial \over \partial  \phi}\right] \ln (p \rho^{-\gamma})=\nonumber \\
&&\qquad{1\over R}{\partial \over \partial R}\left(R \kappa T^{5/2}{\partial T\over \partial R}\right)+ {\partial
    \over \partial z}\left(\kappa T^{5/2}{\partial T\over \partial
      z}\right)+ {1\over R}{\partial \over \partial
    \phi}\left({\kappa T^{5/2}\over R}{\partial T\over \partial
      \phi}\right)-\left({\rho\over \mu
      \mp}\right)^2\Lambda(T).
\label{eq:hydrocyle}
\end{eqnarray}
The unperturbed atmosphere is assumed to be axisymmetric and close to
hydrostatic and thermal equilibrium, in the sense that the system is
approximately in a steady state over the time scales of interest even
in the presence of radiative cooling and thermal conduction.  Thus,
the unperturbed fluid can be described by the time-independent
axisymmetric pressure $\pzero$, density $\rhozero$, temperature
$\Tzero$ and velocity field $\vvzero=(\vzeroR,\vzeroz,\vzerophi)$,
which satisfy equations~(\ref{eq:hydrocylc}-\ref{eq:hydrocyle}) with
vanishing partial derivatives with respect to $t$.  The velocity
components $\vzeroR$ and $\vzeroz$ are allowed to be non-null, because
an inflow of gas can occur if cooling is not perfectly balanced by
thermal conduction. Such a steady-state configuration is expected to
be a reasonable approximation for the corona sufficiently far from the
disc, where we are interested in studying the thermal instability. The
model is not intended to apply near the disc and close to the galactic
centre, where the interactions of the coronal gas with the
star-forming disc and the central super-massive black hole are
important.

In the most general case the atmosphere is assumed to be
differentially rotating, with angular velocity
$\Omega\equiv\vzerophi/R$ depending on both $R$ and $z$, but in the
analysis it is convenient to distinguish cases with
$\partial\Omega/\partial z=0$ from cases with $\partial\Omega/\partial
z\neq0$. The Poincar\'e-Wavre theorem \citep[e.g.][]{Tas78} states
that the surfaces of constant pressure and constant density coincide
if and only if $\partial\Omega/\partial z=0$: as a consequence,
distributions with $\partial\Omega/\partial z=0$ are said {\it
  barotropic} (pressure is a function of only density), while
distributions with $\partial\Omega/\partial z\neq0$ are said {\it
  baroclinic} (pressure is not a function of only density).


To address the question of the thermal stability of the rotating
corona, we study the behaviour of the system in the presence of small
(linear) thermal perturbations. These perturbations are not expected
to have particular symmetries in a real system, so the relevant case
is that of non-axisymmetric perturbations in the presence of
differential rotation.  It is well known that in a differentially
rotating system such a problem is complicated by the effect of the
shearing background on the perturbations~\citep{Cow51,Gol65,Lyn67}.
The analysis is much simpler in the case of axisymmetric perturbations
or uniform rotation: though these assumptions are not expected to
apply in general to galactic coronae, exploring these cases is useful
to understand the behaviour of the more realistic, but far more
complex case.  Therefore, before addressing the full problem of
non-axisymmetric perturbations in a differentially rotating corona, we
will consider the cases of axisymmetric perturbations in a
differentially rotating atmosphere and of non-axisymmetric
perturbations under the assumption of uniform rotation.

\begin{table}
\caption{List of commonly used symbols.
  \label{tab:list}}
\begin{tabular}{ll}

$\ApR\equiv(\pd \pzero/\pd R) /\pzero$ & Inverse of the pressure scale-length\\

$\Apz\equiv(\partial \pzero/\pd z) /\pzero$& Inverse of the pressure scale-height\\

$\ArhoR\equiv(\pd \rhozero/\pd R) /\rhozero$&  Inverse of the density scale-length \\

$\Arhoz\equiv(\partial \rhozero/\pd z) /\rhozero$& Inverse of the density scale-height\\

$\czero\equiv (\pzero/\rhozero)^{1/2}$, $\czerotil\equiv\czero/\Omega R$& Isothermal sound speed, normalized isothermal sound speed\\

$\Dm\equiv {(\kR/\kz)}{\partial/\partial z}-{\partial/\partial R}$& Derivative along surfaces of constant wave-phase\\


$\kv=(\kR,\kz,\kphi)$& Perturbation wave-vector\\

$k=|\kv|=(\kR^2+\kz^2+\kphi^2)^{1/2}$& Perturbation wave-number\\

$\kRtil\equiv\kR/\kphi$, $\kztil\equiv\kz/\kphi$, $\ktil\equiv k/\kphi$& Normalized wave-vector components and wave-number\\

$n\equiv-\i\omegahat$, $\ntilde\equiv n/|\omegad|$& Doppler-shifted perturbation frequency\\

$p$& Pressure perturbation throughout the paper (pressure in
Section~2)\\

$\pzero=\pzero(R,z)$& Unperturbed pressure\\

$\szero\equiv \ln \pzero \rhozero^{-\gamma}$& Unperturbed specific entropy\\

$T$& Temperature perturbation throughout the paper (temperature in
Section~2) \\

$\Tzero=\Tzero(R,z)$& Unperturbed temperature\\

$\vvzero=\vvzero(R,z)=(\vzeroR,\vzeroz,\vzerophi)$& Unperturbed velocity \\

$\vR$, $\vz$, $\vphi$ & Components of the velocity perturbation throughout the paper (of the velocity in Section~2) \\

$\vRtil\equiv\vR/\Omega R$, $\vztil\equiv\vz/\Omega R$ & Normalized components of the velocity perturbation\\


$\gamma=\frac{5}{3}$& Ratio of principal specific heats\\

$\gammaprime\equiv{\d\ln \pzero/ \d \ln \rhozero}$ & Local polytropic index of barotropic distributions.\\

$\GammapR\equiv{\pd \ln \pzero / \pd \ln R}$,
  $\Gammapz\equiv({R/z}){\pd \ln \pzero / \pd \ln |z|}$ & Local $R$ and
  $z$ logarithmic slopes of the unperturbed pressure\\

$\GammarhoR\equiv{\pd \ln \rhozero / \pd \ln R}$,
  $\Gammarhoz\equiv({R/z}){\pd \ln \rhozero / \pd \ln |z|}$ & Local $R$ and
  $z$ logarithmic slopes of the unperturbed density\\

$\GammaOmegaR\equiv{\pd \ln \Omega / \pd \ln R}$,
  $\GammaOmegaz\equiv({R/z}){\pd \ln \Omega / \pd \ln |z|}$ & Local $R$ and
  $z$ logarithmic slopes of the angular velocity\\

$\kappa$& Thermal conductivity\\

$\Lambda$& Radiative cooling function\\

$\mu$ & Mean gas particle mass in units of the proton mass $\mp$\\

$\rho$& Density perturbation throughout the paper (density in
Section~2)\\

$\rhozero=\rhozero(R,z)$& Unperturbed density\\

$\rhotil\equiv \rho/\rhozero$& Normalized density perturbation\\

$\tau\equiv t \Omega$ & Normalized time\\

$\omega$& Perturbation frequency\\

$\omegahat\equiv\omega-\kv\cdot\vvzero$& Doppler-shifted perturbation frequency\\

$\omegaBV$, $\omegaBVtil\equiv\omegaBV/|\omegad|$ &
Brunt-V\"ais\"al\"a (or buoyancy) frequency (equations~\ref{eq:omegaBV} and \ref{eq:omegaBVbarotr})\\

$\omegarot$, $\omegarottil\equiv\omegarot/|\omegad|$ &
Characteristic frequency associated with rotation (equations~\ref{eq:omegarot} and \ref{eq:omegarotbarotr})\\

$\omegath$, $\omegathtil\equiv\omegath/\Omega$ & Thermal-instability frequency (equation~\ref{eq:omegath})\\

$\omegac$, $\omegactil\equiv (\omegac \kphi^2)/(\Omega k^2)$& Thermal-conduction frequency (equation~\ref{eq:omegac})\\

$\omegad\equiv\omegath+\omegac$, $\omegadtil\equiv\omegad/\Omega$ & Characteristic frequency of dissipative processes\\

$\Omega=\vzerophi/R$& Angular velocity\\

$\OmegaR\equiv\partial(\Omega R)/\pd R$, $\Omegaz\equiv\partial(\Omega R)/\pd z$ & Local $R$ and $z$ derivatives of $\vzerophi=\Omega R$ \\
\end{tabular}
\end{table}

\section{Thermal-stability analysis: axisymmetric perturbations}
\label{sec:axisymm}

\subsection{Derivation of the dispersion relation}

Here the fluid is assumed to rotate differentially with
$\Omega=\Omega(R,z)$ and that the perturbations are axisymmetric.
Linearizing the hydrodynamic
equations~(\ref{eq:hydrocylc}-\ref{eq:hydrocyle}) with perturbations
of the form $\Fzero+F\exp(-\i\omega t+\i\kR R + \i\kz z)$ (where
$\Fzero$ is the unperturbed quantity and $|F|\ll|\Fzero|$), in the
limit of short-wavelength, low-frequency perturbations we get
\begin{eqnarray}
 &&-\i\omegahat \rho + \i\kR\vR\rhozero+\i\kz\vz\rhozero=0,\label{eq:axc}\\
 && -\i\omegahat\vR\rhozero=-\i\kR p+\ApR\czero^2\rho+2\Omega\vphi\rhozero,\label{eq:axm1}\\
 && -\i\omegahat\vz\rhozero=-\i\kz p+\Apz\czero^2\rho,\label{eq:axm2}\\
 && -\i\omegahat\vphi\rhozero+\vR\rhozero{\OmegaR}+\vz\rhozero{\Omegaz}= - \rhozero \vR\Omega,\label{eq:axm3}\\
&&{\Tzero\over T\gamma}\left[-\i\omegahat{p\over\pzero}+\i\gamma\omegahat{\rho\over\rhozero}
+\vR\left(\ApR-\gamma\ArhoR\right)+\vz\left(\Apz-\gamma\Arhoz\right)\right]=-(\omegac+\omegath),\label{eq:axe}
\end{eqnarray}
where
$\omegahat\equiv\omega-\kv\cdot\vvzero=\omega-(\kR\vzeroR+\kz\vzeroz)$
is the Doppler-shifted frequency, $\czero^2\equiv \pzero/\rhozero$ is
the isothermal sound speed squared, $\ApR\equiv(\pd \pzero/\pd R)
/\pzero$ and $\Apz\equiv(\partial \pzero/\pd z) /\pzero$ are the
inverse of the pressure scale-length and scale-height,
respectively. The following frequencies have also been defined:
$\OmegaR\equiv\partial(\Omega R)/\pd R$, $\Omegaz\equiv\partial
(\Omega R)/\pd z$, the thermal-conduction frequency
\begin{equation}
\omegac\equiv\left({\gamma-1\over\gamma}\right){k^2 \kappa \Tzero^{7/2}\over \pzero},\label{eq:omegac}
\end{equation}
and the thermal-instability frequency
\begin{equation}
\omegath\equiv-\left({\gamma-1\over\gamma}\right){\rhozero^2\Lambda(\Tzero)\over
\pzero(\mu \mp)^2}\left[2-{\d\ln \Lambda (\Tzero) \over \d \ln
\Tzero}\right].\label{eq:omegath}
\end{equation}
In terms of the defined quantities, the assumption of short-wavelength
perturbations gives $|\kR|,|\kz|\gg|\ArhoR|,|\Arhoz|,|\Apz|,|\ApR|$,
and $\Omega^2,\OmegaR^2,\Omegaz^2\ll \czero^2 k^2$, while the
assumption of low-frequency perturbations gives $\omega^2 \ll \czero^2
k^2$. In deriving the energy equation~(\ref{eq:axe}) we used $\rhozero
T\simeq -\Tzero\rho$, because it can be shown that in the considered
limit $p/\pzero\ll \rho/\rhozero$ (see Appendix~\ref{app:isob}).  The
system of equations~(\ref{eq:axc}-\ref{eq:axe}) can be reduced to the
following dispersion relation\footnote{The same dispersion relation is
  obtained by working from the beginning in the Boussinesq
  approximation, i.e. neglecting the term $-\i\omegahat \rho$ in
  equation~(\ref{eq:axc}) and the term $-\i\omegahat p/\pzero$ in
  equation~(\ref{eq:axe}).} for $n\equiv-\i\omegahat$:
\begin{equation}
n^3 +n^2\omegad +(\omegaBVsq+\omegarotsq)n+\omegarotsq\omegad=0,
\label{eq:disp}
\end{equation}
where $\omegad\equiv\omegath+\omegac$ is the characteristic frequency
of dissipative processes,
\begin{equation}
\omegarotsq\equiv-\kzksq{1\over R^3} \Dm({R^4 \Omega^2})
\label{eq:omegarot}
\end{equation}
is the differential rotation term,
\begin{equation}
\omegaBVsq\equiv-\kzksq{\Dm \pzero \over \rhozero\gamma}\Dm \szero
\label{eq:omegaBV}
\end{equation}
is the buoyancy term, $\szero\equiv \ln \pzero \rhozero^{-\gamma}$ is
the unperturbed specific entropy and, following \citet{Bal95}, we
introduced the differential operator
\begin{equation}
\Dm\equiv {\kR\over\kz}{\partial\over\partial z}-{\partial\over\partial R},
\label{eq:Dm}
\end{equation}
which can be seen as taking derivatives along surfaces of constant
wave phase.  It may be useful to note that, by definition,
$\omegadsq\geq0$, while $\omegaBVsq$ and $\omegarotsq$ are not
necessarily non-negative, and that $n$ is defined so that the
perturbation evolves in time as $\propto \exp(nt)$, thus the stable
modes are those with ${\rm Re}(n)\leq0$.

\subsection{Limiting cases}
\label{sec:lim}

Before analyzing the dispersion relation~(\ref{eq:disp}) in the
general case, it is useful to discuss some limiting cases, which can
be obtained when one or two among the three characteristic frequencies
$\omegaBV$, $\omegarot$ and $\omegad$ are zero.  Let us start from the
simplest cases, in which only one of these is non-null.
\begin{enumerate}
\item {\it Case with $\omegad=0$ and $\omegarot=0$}.  In this case
  there is no dissipation and the fluid is either non-rotating
  ($\Omega=0$) or rotating differentially with vanishing gradient of
  the specific angular momentum [$\d({\Omega R^2 })/\d R=0$ when
    $\Omega=\Omega(R)$]. From equation~(\ref{eq:disp}) we have the
  dispersion relation
\begin{equation}
n^2 =-\omegaBVsq, 
\end{equation}
so we have stability if the square of the Brunt-V\"ais\"al\"a
frequency $\omegaBVsq>0$. When $\Omega=\Omega(R)$ the
Brunt-V\"ais\"al\"a frequency squared con be written
\begin{equation}
\omegaBVsq=\kzksq{\czero^2\Apz^2\over \gamma}\left({\gamma\over \gammaprime}-1\right)\left({\kR\over \kz}-{\ApR\over\Apz}\right)^2 ,
\label{eq:omegaBVbarotr}
\end{equation}
where, using the fact that $\pzero$ is a function of only $\rhozero$
(because the distribution is barotropic; see Section~\ref{sec:eq}), we
defined
\begin{equation}
\gammaprime\equiv{\d\ln \pzero\over \d \ln \rhozero},
\label{eq:gammaprime}
\end{equation}
which can be considered a local polytropic index.
It follows that, independently of the value of $\kR/\kz$, the
condition for convective stability is $\gammaprime<\gamma$,
i.e. Schwarzschild's criterion \citep[e.g.][]{Tas78}.

\item {\it Case with $\omegaBV=0$ and $\omegarot=0$}. Here we assume
  that the characteristic frequencies associated with rotation and
  buoyancy are null. In the barotropic case these conditions are met
  when $\gammaprime=\gamma$ (i.e. the radial gradient of the specific
  entropy is zero) and ${\d({\Omega R^2})/\d R}=0$ (i.e. the radial
  gradient of the specific angular momentum is zero).  The dispersion
  relation is
\begin{equation}
n=-\omegad,
\end{equation}
so we have thermal instability if $\omegad<0$, which is just Field's
instability criterion \citep{Fie65}. From the definition of
$\omegad\equiv\omegath+\omegac$, it is clear that the condition for
thermal instability is that the growth rate of the thermal
perturbation ($|\omegath|$) must be faster than conductive damping (we
recall that $\omegac \geq 0$, while typically $\omegath<0$).  For
fixed unperturbed gas temperature $\Tzero$ and pressure $\pzero$,
$\omegac$ increases for increasing perturbation wave-number $k$
(equation~\ref{eq:omegac}), while $\omegath$ is independent of $k$
(equation~\ref{eq:omegath}), so there is a critical perturbation
wavelength such that $\omegad<0$ for longer wavelengths and
$\omegad>0$ for shorter wavelengths.

\item {\it Case with $\omegaBV=0$ and $\omegad=0$}. In the absence of
  buoyancy and dissipation we obtain the dispersion relation
\begin{equation}
n^2=-\omegarotsq,
\end{equation}
so the stability criterion is $\omegarotsq>0$.  The value of
$\omegarotsq$ depends on the ratio $\kR/\kz$: $\omegarotsq$ is
positive for all values of $\kR/\kz$ if and only if $\partial \Omega
/\partial z=0$ and $\d({\Omega R^2})/\d R>0$, i.e. the specific
angular momentum must increase outwards: this is just Rayleigh's
stability criterion \citep{Chandra61}.
\end{enumerate}
In the following three limiting cases of the dispersion
relation~(\ref{eq:disp}) only one among $\omegaBV$, $\omegarot$ and
$\omegad$ is null.
\begin{enumerate}
\item {\it Case with $\omegaBV=0$}. In the absence of buoyancy, but
  for $\omegarot\neq0$ and $\omegad\neq0$, the dispersion relation is
\begin{equation}
(n^2+\omegarotsq)(\omegad+n)=0,
\end{equation}
which is just a combination of the criteria obtained in the points ii)
and iii) above, so (when $\omegaBV=0$) the presence of a gradient of
the specific angular momentum does not modify the thermal-instability
criterion in an interesting way. Specifically, when $\omegad<0$ the
medium is thermally unstable, independent of the presence and
properties of rotation, while rotation can destabilize an otherwise
thermally stable medium ($\omegad>0$) if $\omegarotsq<0$. Thus, the
condition for stability is $\omegad>0$ and $\omegarotsq>0$, which
holds for all values of $\kR/\kz$ if and only if $\partial \Omega
/\partial z=0$ and $\d({\Omega R^2})/\d R>0$.

\item {\it Case with $\omegad=0$}. In the absence of dissipation, but
  for $\omegarot\neq0$ and $\omegaBV\neq0$, one obtains the dispersion
  relation
\begin{equation}
n^2 =-(\omegaBVsq+\omegarotsq),
\end{equation}
which leads to the well-known convective stability criterion for a
rotating, stratified fluid $\omegaBVsq+\omegarotsq>0$, showing the
stabilizing effect of rotation against convection.  The inequality
$\omegaBVsq+\omegarotsq>0$ is verified for all values of $\kR/\kz$ if
and only if
\begin{equation}
-{1\over \gamma \rhozero}\grad \pzero \cdot \grad \szero +{1\over R^3} {\partial R^4\Omega^2 \over \partial R} > 0
\end{equation}
and 
\begin{equation}
-{\partial \pzero \over \partial z}\left({\partial R^4\Omega^2 \over \partial R} {\partial \szero \over \partial z}-{\partial R^4\Omega^2 \over \partial z} {\partial \szero \over \partial R}\right)> 0,
\end{equation}
which is the classical Solberg-H{\o}iland criterion
\citep[see][]{Sol36,Hoi41,Gol67,Tas78,Bal95}.

\item {\it Case with $\omegarot=0$}.  When the fluid does not rotate,
  or, more generally, has a vanishing gradient of the specific angular
  momentum [$\d({\Omega R^2 })/\d R=0$ in the barotropic case], but
  $\omegad\neq0$ and $\omegaBV\neq 0$, the dispersion relation is
\begin{equation}
n^2 + \omegad n+\omegaBVsq=0.
\label{eq:dispnorot}
\end{equation}
If the system is spherically symmetric, equation~(\ref{eq:dispnorot})
reduces to the dispersion relation derived in MRB (see also BNF). When
$\omegad>0$ we have stability (damping by thermal conduction) if
$\omegaBVsq>0$, while we have convective instability if
$\omegaBVsq<0$.  When $\omegad<0$ we have overstability if
$\omegaBVsqtil>\frac{1}{4} $, thermal instability if
$0<\omegaBVsqtil<\frac{1}{4}$ and convective instability if
$\omegaBVsqtil<0$.
\end{enumerate}

\subsection{Analysis of the general form of the dispersion relation}

Here we consider the dispersion relation~(\ref{eq:disp}) in the
general case in which all coefficients are non-null.  Dividing by
$|\omegad|^3$ we get
\begin{equation}
\ntilde^3 +\ntilde^2 +(\omegaBVsqtil+\omegarotsqtil)\ntilde+\omegarotsqtil=0,\qquad {\rm if}\qquad \omegad>0,
\label{eqdispplus}
\end{equation}
and
\begin{equation}
\ntilde^3 -\ntilde^2 +(\omegaBVsqtil+\omegarotsqtil)\ntilde-\omegarotsqtil=0,\qquad {\rm if}\qquad \omegad<0,
\label{eqdispmin}
\end{equation}
where the dimensionless quantities $\ntilde\equiv n/|\omegad|$,
$\omegaBVsqtil\equiv \omegaBVsq/\omegadsq$ and $\omegarotsqtil \equiv
\omegarotsq/\omegadsq$ have been introduced.  In both cases the
discriminant of the cubic equation is
\begin{equation}
\Delta=-27 x^2 + (36 y +4) x - 4 y (y^2+2y+1),
\end{equation}
where $x=\omegaBVsqtil$ and $y=\omegaBVsqtil+\omegarotsqtil$.

Let us discuss first the case $\omegad>0$ (equation~\ref{eqdispplus}
and diagram in the left-hand panel of Fig.~\ref{fig:axisymm}).
Applying the Routh-Hurwitz theorem (see Appendix~\ref{app:cubic}) we
have that the real parts of all roots are negative (stability) if and
only if $\omegaBVsqtil>0$ and $\omegarotsqtil>0$. Thus the first
quadrant of the diagram in the left-hand panel of
Fig.~\ref{fig:axisymm} is a locus of stable configurations.
Configurations in the other three quadrants can be either unstable or
overstable. The blue curves in the diagrams correspond to discriminant
$\Delta=0$. In the bottom-left area defined by these curves
$\Delta>0$, thus we have three real roots (at least one negative), so
the corresponding configurations are unstable. In the other regions
$\Delta<0$, so we have one real root ($\ntilde_1$) and two complex
conjugate roots ($\ntilde_2$ and $\ntilde_3$).  The sign of the real
root $\ntilde_1$ can be determined because we know that
$\ntilde_1\ntilde_2\ntilde_3=-\omegarotsqtil$ (see
Appendix~\ref{app:cubic}), and $\ntilde_2\ntilde_3$ is obviously
positive. Thus we have instability in the upper-left region and
overstability in the bottom-right region.  In summary, the condition
for stability against axisymmetric perturbations is $\omegaBVsq>0$ and
$\omegarotsq>0$, which holds for all values of $\kR/\kz$ if $\partial
\Omega /\partial z=0$ [thus $\Omega=\Omega(R)$ and
  $\pzero=\pzero(\rhozero)$], $\d({\Omega R^2})/\d R>0$ and $\d\ln
\pzero / \d \ln \rhozero<\gamma$. We note that the analysis of the
case $\omegad>0$ was carried out also by \cite{Lif93} in the context
of the study of the stability of rotating stars in the presence of
radiative diffusion~\citep[see also][]{Gol67,Sun74b,Sun75,Bal01}.

Let us move to the case $\omegad<0$ (equation~\ref{eqdispmin} and
diagram in the right-hand panel of Fig.~\ref{fig:axisymm}). The same
line of reasoning as for the case $\omegad>0$ leads to the following
conclusions.  Given that the coefficient of $\ntilde^2$ is negative,
we know from the Routh-Hurwitz theorem that in no case all the real
parts of the roots are negative.  In other words, all configurations
will be either unstable or overstable.  The blue curves in the diagram
in the right-hand panel of Fig.~\ref{fig:axisymm} correspond to
discriminant $\Delta=0$.  In the bottom-left area defined by these
curves $\Delta>0$, thus we have three real roots (at least one
negative), so the corresponding configurations are unstable.  In the
other regions $\Delta<0$, so we have one real root and two complex
conjugate roots: in this case the sign of the real root is the same as
the sign of $\omegarotsqtil$.  Thus we have overstability in the
upper-left region and instability elsewhere.  It follows that a
necessary condition for overstability is $\omegaBVsq>0$ and
$\omegarotsq<0$, which holds for all values of $\kR/\kz$ if $\partial
\Omega /\partial z=0$ [thus $\Omega=\Omega(R)$ and
  $\pzero=\pzero(\rhozero)$], $\d({\Omega R^2})/\d R<0$ and
$\gammaprime<\gamma$. Note that this condition is not sufficient,
because for positive, but small enough values of $\omegaBVsq$
instability replaces overstability even if $\omegarotsq<0$ (see
top-left quadrant of the right-hand panel of
Fig.~\ref{fig:axisymm}). The non-rotating case (briefly discussed at
the end Section~\ref{sec:lim}) is obtained from the diagram at
$\omegarotsqtil=0$ (overstability if $\omegaBVsqtil>\frac{1}{4}$, and
instability at lower values of $\omegaBVsqtil$; see MRB).

\begin{figure}
\centerline{\psfig{file=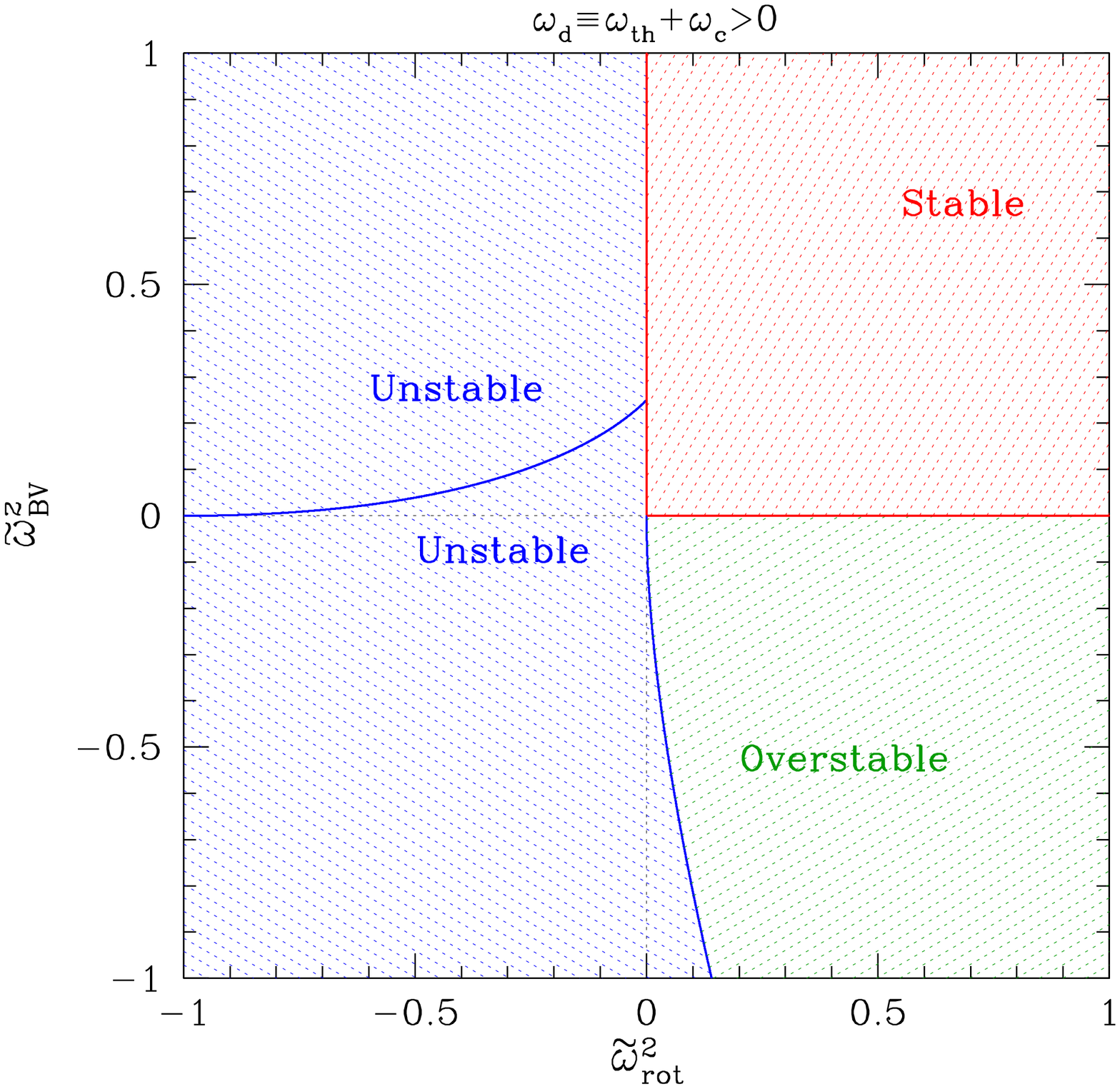,width=0.5\hsize}
\psfig{file=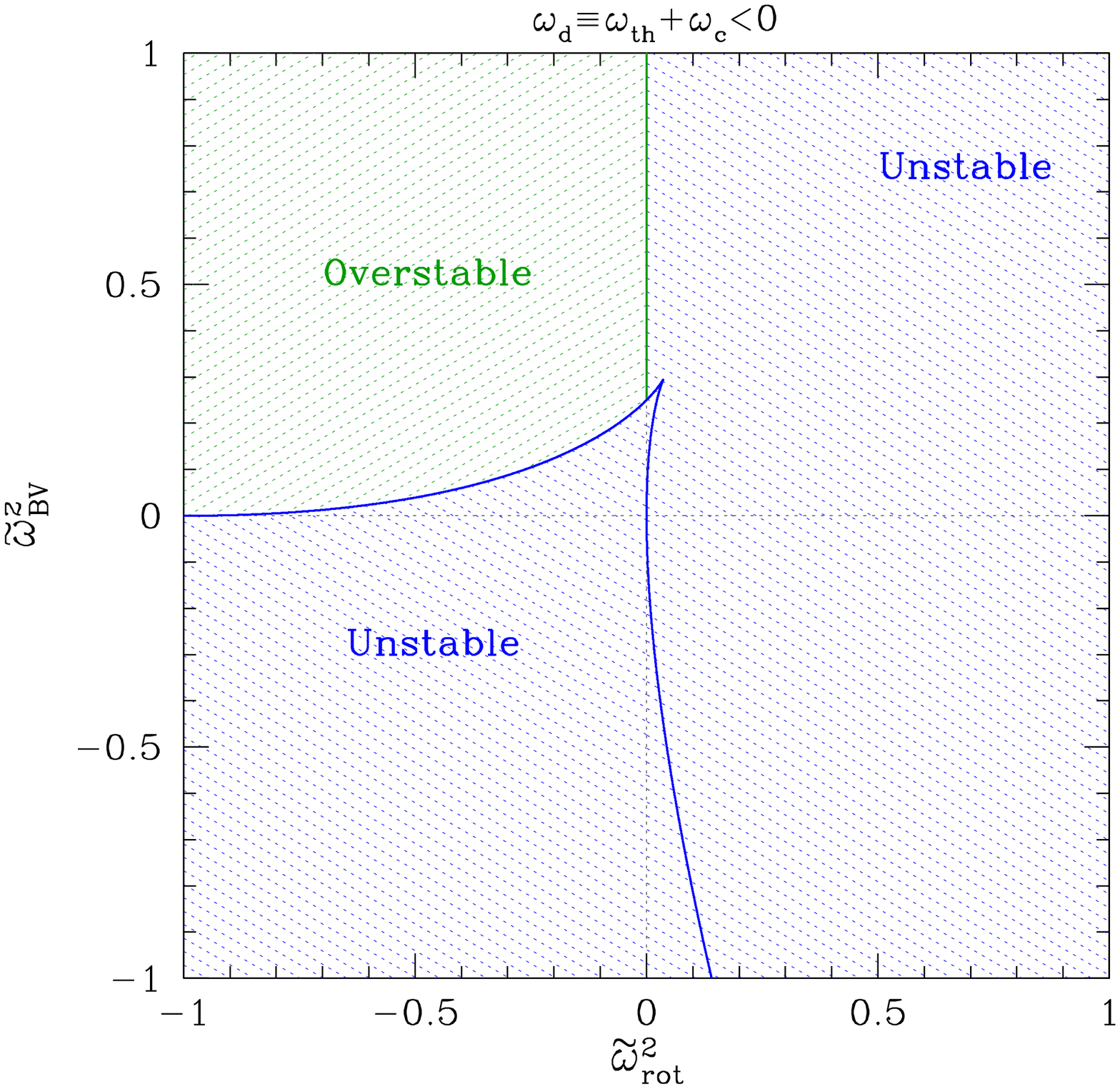,width=0.5\hsize}}
\caption{Domain of stability (red), overstability (green) and
  instability (blue) against axisymmetric perturbations in the plane
  $\omegaBVsqtil\equiv\omegaBVsq/|\omegad|^2$ versus
  $\omegarotsqtil\equiv\omegarotsq/|\omegad|^2$, for a differentially
  rotating, stratified fluid, when $\omegad>0$ (left-hand panel) and
  $\omegad<0$ (right-hand panel).  $\omegarotsq$ and $\omegaBVsq$ are
  given, respectively by equations~(\ref{eq:omegarot})
  and~(\ref{eq:omegaBV}).}
\label{fig:axisymm}
\end{figure}

\begin{figure}
\centerline{\psfig{file=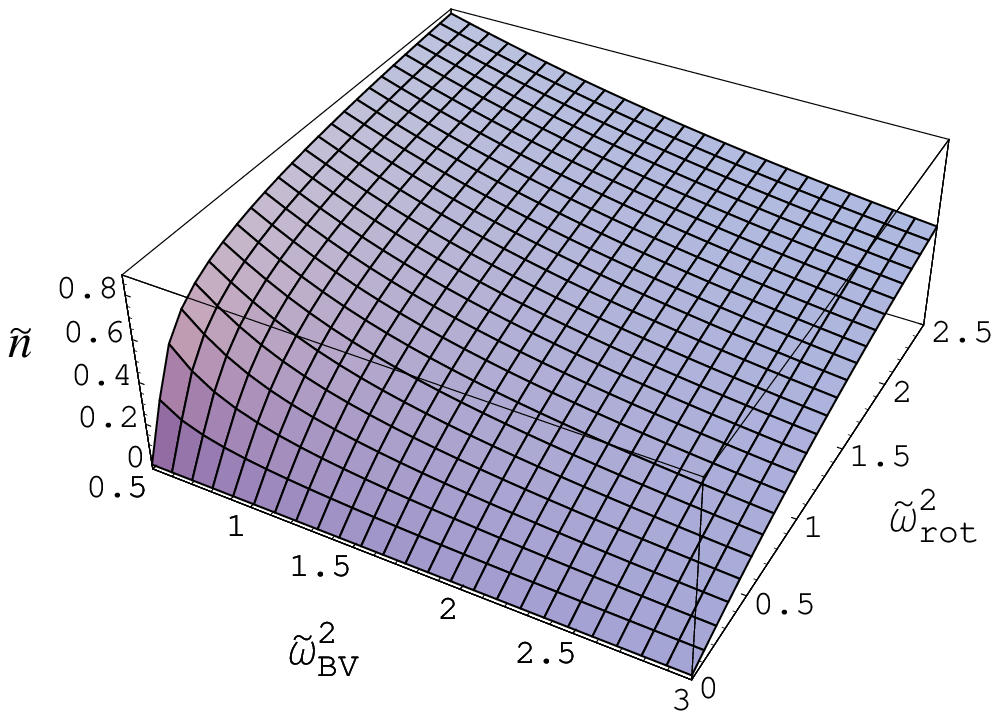,width=0.5\hsize}}
\caption{Growth rate $\ntilde$ of axisymmetric perturbations as a
  function of $\omegarotsqtil$ and $\omegaBVsqtil$ for the unstable
  mode in the first quadrant ($\omegaBVsq>0$, $\omegarotsq>0$) of the
  right-hand panel ($\omegad<0$) of Fig.~\ref{fig:axisymm}.}
\label{fig:rate}
\end{figure}

\subsection{Discussion}

The above results indicate that the signs of $\omegaBVsq$ and
$\omegarotsq$ are crucial for the thermal-stability properties of a
configuration.  Both $\omegaBVsq$ and $\omegarotsq$ depend not only on
the properties of the unperturbed distribution, but also on the
wave-vector of the perturbations through the coefficient $\kz^2/k^2$,
and through $\kR/\kz$, which appears in the differential operator
$\Dm$ defined in equation~(\ref{eq:Dm}).  A general result, valid for
both positive and negative $\omegad$, is that if $\Omega$ depends on
$z$ the system is unstable, in the sense that it is always possible to
choose $\kR/\kz$ such that we are in the instability region of
parameters.  To look for stable or overstable configurations we must
consider barotropic distributions, characterized by $\Omega=\Omega(R)$
and $\pzero=\pzero(\rhozero)$.  In this case $\omegarotsq$ is given by
\begin{equation}
\omegarotsq =\kzksq{1\over R^3} {\d \over \d R}({R^4 \Omega^2}),
\label{eq:omegarotbarotr}
\end{equation}
so it reduces to $(\kz/k)^2$ times Rayleigh's discriminant
\citep{Chandra61}. The condition for $\omegarotsq>0$ is
\begin{equation}
 {\d \ln \Omega \over \d \ln R}>-2,
\label{eq:ray}
\end{equation}
which is Rayleigh's stability criterion, requiring that the specific
angular momentum must increase outwards.  In addition, for barotropes
the Brunt-V\"ais\"al\"a frequency is given by
equation~(\ref{eq:omegaBVbarotr}), so the condition for $\omegaBVsq>0$
is Schwarzschild's criterion
\begin{equation}
\gammaprime<\gamma.
\label{eq:sch}
\end{equation} 

In most applications we are interested in the case of convectively
stable systems (satisfying condition~\ref{eq:sch}) with
outward-increasing specific angular momentum~(satisfying
condition~\ref{eq:ray}).  Coming back to the diagrams in
Fig.~\ref{fig:axisymm} we conclude that in the presence of cooling and
thermal conduction, and assuming axisymmetric perturbations, these
systems are stable if $\omegad>0$ (damping by thermal conduction),
while are unstable if $\omegad<0$, independent of the value of the
ratio $\omegaBVsq/\omegad^2$ and on of the value of $\omegarotsq>0$.
This means that in the case $\omegad<0$ (i.e., when the perturbation
wavelength is short enough that the thermal-instability frequency is
higher than the conductive-damping frequency) even arbitrarily slow
rotation changes qualitatively the behaviour of the system
transforming overstability (found in the absence of rotation; MRB)
into instability.  Physically, this different behaviour is due to the
fact that, when the perturbations are axisymmetric, conservation of
angular momentum tends to inhibit convection \citep{Cow51} and thus to
prevent overdense regions from falling inward far enough to oscillate
around the radius proper to their particular value of specific
entropy.  Note that the transition between overstability and
instability is not abrupt, because the growth rate of the unstable
mode goes to zero when $\omegarotsq\to0$ (for
$\omegaBVsqtil>\frac{1}{4}$).  This is apparent from
Fig.~\ref{fig:rate}, plotting the instability growth rate $\ntilde$,
which is the real root obtained by solving equation~(\ref{eqdispmin})
for $\omegarotsqtil>0$ and $\omegaBVsqtil>0$, as a function of
$\omegarotsqtil$ and $\omegaBVsqtil$.  The growth rate, at fixed
$\omegaBVsq$, increases gradually from 0 for $\omegarotsqtil$
increasing from 0; the larger $\omegaBVsqtil$ the larger must be
$\omegarotsqtil$ to give substantial growth rates.

\section{Thermal-stability analysis: non-axisymmetric perturbations}
\label{sec:nonaxisymm}

In the present Section we move to the next step of the stability
analysis, which is the study of non-axisymmetric
perturbations. Formally, it would be sufficient to perform the
non-axisymmetric perturbation analysis for those configurations that
are stable or overstable against non-axisymmetric perturbations, which
might turn out to be unstable if more general perturbations are
considered. Nevertheless, we will perform the non-axisymmetric
perturbation analysis also for configurations found unstable against
non-axisymmetric perturbations, because it is interesting to verify
whether axisymmetry is a necessary condition for a perturbation to
grow. 

There are reasons to expect that non-axisymmetric thermal
perturbations behave differently from axisymmetric ones. \citet{Cow51}
showed that rotation has a stabilizing effect against convection for
axisymmetric perturbations (a consequence of angular momentum
conservation), but not necessarily for non-axisymmetric perturbations:
in the latter case the stabilizing effect of rotation is less and it
vanishes for specific perturbations. This suggests that rotating
galactic coronae (in the absence of heat conduction) might be
thermally unstable against axisymmetric perturbations, but not against
non-axisymmetric perturbations, because buoyancy should prevail,
provided the gas entropy increases outwards \citep[see also][]{Nip10}.

We now address quantitatively the problem of stability against
non-axisymmetric perturbations.  As mentioned above, in this case the
analysis is rather complicated if the corona rotates differentially
\citep{Gol65}.  Therefore, we first try to get an estimate of the
effect of the deviation from axisymmetry of the perturbations by
considering the simpler case of non-axisymmetric perturbations in a
uniformly rotating corona (Section~\ref{sec:unif}), and we defer the
treatment of the differentially rotating case to
Section~\ref{sec:diff}.

\subsection{Uniform rotation}
\label{sec:unif}

\begin{figure}
\centerline{\psfig{file=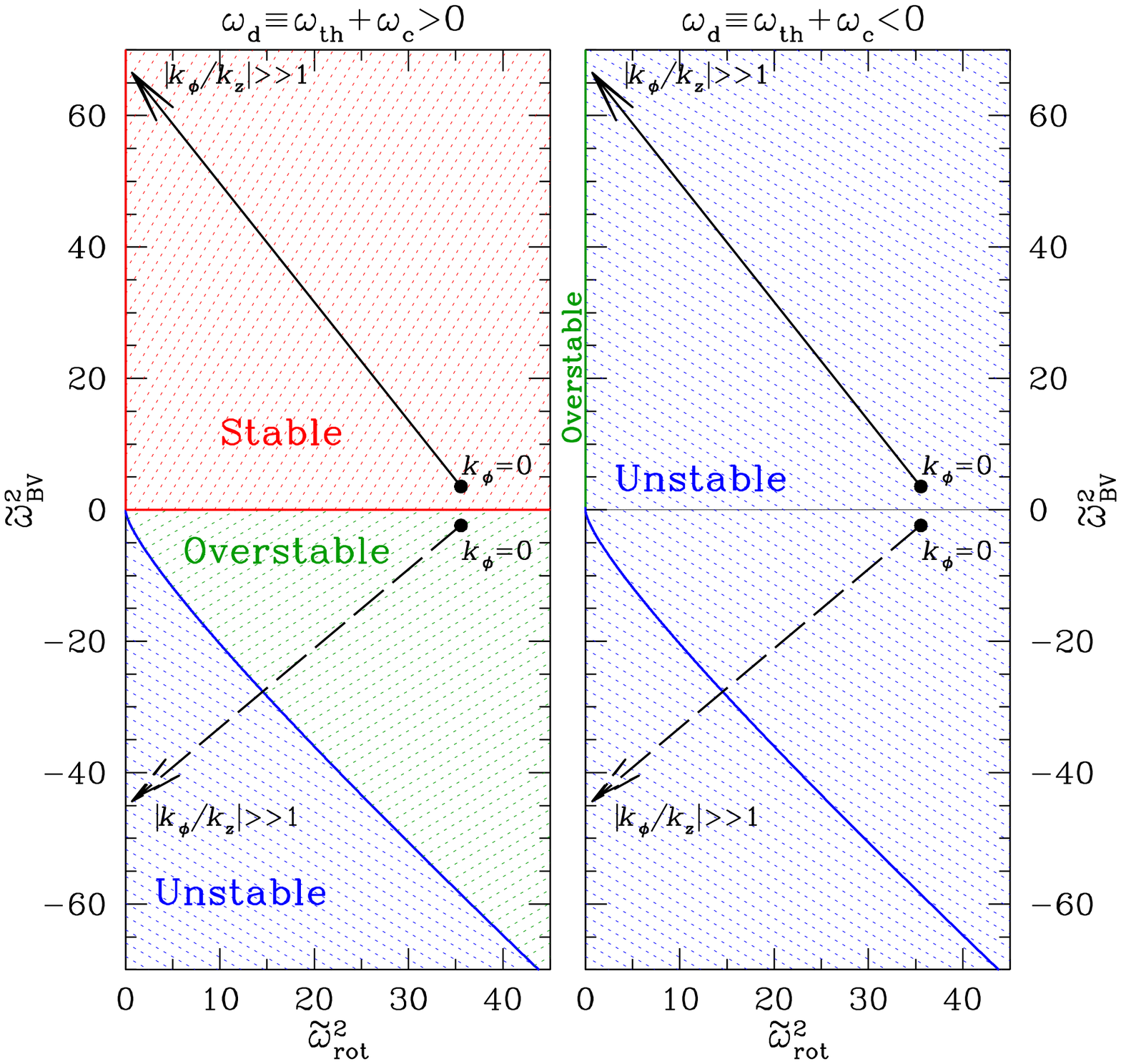,width=0.5\hsize}}
\caption{Domain of stability (red), overstability (green) and
  instability (blue) against non-axisymmetric perturbations in the
  plane $\omegaBVsqtil\equiv\omegaBVsq/|\omegad|^2$
  (equation~\ref{eq:nonaxomegaBVtil}) versus
  $\omegarotsqtil=\omegarotsq/|\omegad|^2$
  (equation~\ref{eq:nonaxomegarottil}), for a uniformly rotating,
  stratified gas, when $\omegad>0$ (left-hand panel) and when
  $\omegad<0$ (right-hand panel).  The arrows show the effect of
  increasing the azimuthal wave-number $\kphi=m/R$ from $\kphi=0$
  (filled circles) to $|\kphi/\kz|\gg 1$, for representative cases with
  $\gammaprime=1$ (solid lines) and $\gammaprime=3$ (long dashed
  line), assuming $\kz/\kR=2$, $|\omegad|/\Omega=0.3$, $\czerotil=2$,
  $\Gammapz=-1.5$, and $\GammapR=-1.25$.}
\label{fig:nonax}
\end{figure}

Linearizing the hydrodynamic
equations~(\ref{eq:hydrocylc}-\ref{eq:hydrocyle}) with perturbations
of the form $\Fzero+F\exp(-\i\omega t+\i\kR R + \i\kz z+\i m \phi)$,
where $\Fzero$ is the unperturbed quantity and $|F|\ll|\Fzero|$, and
assuming angular velocity $\Omega$ independent of position, we get:
\begin{eqnarray}
 &&-\i\omegahat \rho + \i\kR\vR\rhozero+\i\kz\vz\rhozero+\i \kphi\vphi\rhozero=0,\\
 && -\i\omegahat\vR\rhozero=-\i\kR p+\ApR\czero^2\rho+2\Omega\vphi\rhozero,\\
 && -\i\omegahat\vz\rhozero=-\i\kz p+\Apz\czero^2\rho,\\
 && -\i\omegahat\vphi\rhozero+2\vR\rhozero \Omega=-\i \kphi p,\\
&&{\Tzero\over T\gamma}\left[-\i\omegahat{p\over\pzero}+\i\gamma\omegahat{\rho\over\rhozero}
+\vR\left(\ApR-\gamma\ArhoR\right)+\vz\left(\Apz-\gamma\Arhoz\right)\right]=-\omegad,
\end{eqnarray}
where now $k^2=\kR^2+\kz^2+\kphi^2$ and
$\omegahat\equiv\omega-\kv\cdot\vvzero=\omega-(\kR\vzeroR+\kz\vzeroz+\kphi\vzerophi)$,
with $\kphi\equiv m/R$. In deriving the equations above we have
assumed, as done in Section~\ref{sec:axisymm}, short wavelengths
($|\kphi|,|\kR|,|\kz|\gg|\ArhoR|,|\Arhoz|,|\ApR|,|\Apz|,\Omega/\czero$),
low frequencies ($\omega^2 \ll \czero^2 k^2$) and $\rhozero T =
-\Tzero\rho$. The system of equations can be reduced to the dispersion
relation
\begin{equation}
n^3 +n^2\omegad +(\omegaBVsq+\omegarotsq)n+\omegarotsq\omegad=0,
\label{eq:nonaxdisp}
\end{equation}
where now
\begin{equation}
\omegaBVsq\equiv-{1 \over \gamma\rhozero}\kzksq\left[{\Dm \pzero}\Dm
  \szero +{\kphi^2\over \kz^2}\left({\partial \pzero \over
    \partial R}{\partial \szero \over \partial R}+{\partial \pzero
    \over \partial z}{\partial \szero \over \partial
    z}\right)\right]=
\kzksq{\czero^2\Apz^2\over \gamma}\left({\gamma\over
  \gammaprime}-1\right)\left[\left({\kR\over \kz}-{\ApR\over\Apz}\right)^2
  +{\kphi^2\over \kz^2}\left(1+{\ApR^2\over \Apz^2}\right) \right]
\label{eq:nonaxomegaBV}
\end{equation}
is the buoyancy term, which for $\kphi=0$ reduces to the
Brunt-V\"ais\"al\"a frequency in equation~(\ref{eq:omegaBVbarotr}),
with $\gammaprime$ given by equation~(\ref{eq:gammaprime}), and
\begin{equation}
\omegarotsq\equiv \kzksq 4\Omega^2
\label{eq:nonaxomegarot}
\end{equation}
is the rotation term, which is just $\omegarotsq$ as defined in
equation~(\ref{eq:omegarot}) for position-independent $\Omega$.  Note
that in deriving the dispersion relation (\ref{eq:nonaxdisp}) we
exploited the relation
\begin{equation}
{\partial \pzero \over \partial z}{\partial \szero \over \partial R}={\partial \pzero \over \partial R}{\partial \szero \over \partial z},
\end{equation}
which holds because the surfaces of constant pressure and constant
specific entropy coincide in fluids rotating with $\Omega$ independent
of $z$ (i.e. with barotropic distributions; see Section~\ref{sec:eq}).

The dispersion relation~(\ref{eq:nonaxdisp}) for non-axisymmetric
perturbations in a uniformly rotating medium is formally identical to
the dispersion relation~(\ref{eq:disp}) for axisymmetric
perturbations. We just note that now $\omegaBVsq$ depends on $\kphi$
(equation~\ref{eq:nonaxomegaBV}) and $\omegarotsq\geq 0$
(equation~\ref{eq:nonaxomegarot}).  When $\omegad=0$ we obtain the
dispersion relation $n^2=-(\omegaBVsq+\omegarotsq)$ (with $\omegaBVsq$
and $\omegarotsq$ given by equations \ref{eq:nonaxomegaBV} and
\ref{eq:nonaxomegarot}), so the necessary and sufficient condition for
convective stability against non-axisymmetric perturbations is
$\omegaBVsq+\omegarotsq>0$, showing the stabilizing effect of uniform
rotation against convection \citep[in accordance with previous
  studies;][]{Cow51,Sun74a,Sun75,Ryu92}.  When $\omegad\neq 0$, the
analysis of the dispersion relation~(\ref{eq:nonaxdisp}) is the same
as that of Section~\ref{sec:axisymm} and leads to the diagrams
reported in Fig.~\ref{fig:nonax}, where now
\begin{eqnarray}
\omegaBVsqtil&\equiv&{\omegaBVsq\over\omegadsq}
=\kzksq{\czerotil^2\Gammapz^2\over \gamma \omegadtil^2}\left({\gamma\over
  \gammaprime}-1\right)\left[\left({\kR\over \kz}-{\GammapR\over\Gammapz}\right)^2
  +{\kphi^2\over \kz^2}\left(1+{\GammapR^2\over \Gammapz^2}\right) \right]
\label{eq:nonaxomegaBVtil}
\end{eqnarray}
and 
\begin{equation}
\omegarotsqtil\equiv{\omegarotsq\over\omegadsq} ={4\over \omegadtil^2}\kzksq,
\label{eq:nonaxomegarottil}
\end{equation}
having introduced the dimensionless quantities
$\czerotil\equiv\czero/\Omega R$, $\omegadtil\equiv\omegad/\Omega$,
\begin{equation}
\GammapR\equiv{\pd \ln \pzero \over \pd \ln R}=R\ApR \quad{\rm and}\quad
  \Gammapz\equiv{R\over z}{\pd \ln \pzero \over \pd \ln
    |z|}=R\Apz.
\label{eq:Gammap}
\end{equation}
For both $\omegad>0$ (left-hand panel in Fig.~\ref{fig:nonax}) and
$\omegad<0$ (right-hand panel in Fig.~\ref{fig:nonax}) the domain of
stability, instability and overstability are the same as in the first
and fourth quadrants of the corresponding diagrams of
Fig.~\ref{fig:axisymm}.

The deviation from axisymmetry of the perturbation is measured by the
ratio $\kphi^2/k^2$, on which both $\omegarotsq$ and $\omegaBVsq$
depend. When the properties of the unperturbed system (in practice,
$\omegadtil$, $\czerotil$, $\GammapR$, $\Gammapz$ and $\gammaprime$)
and the ratio of the $R$ and $z$ wave-numbers $\kR/\kz$ are fixed, the
position in the $\omegarotsqtil$-$\omegaBVsqtil$ plane depends only on
$\kphi^2/k^2$. The sign of $\omegaBVsq$ depends only on $\gammaprime$
(equation~\ref{eq:nonaxomegaBV}), so the quadrant in which a system is
located does not change by varying the ratio $\kphi^2/k^2$ at fixed
$\gammaprime$.  Therefore, when $\omegad>0$ {\it a uniformly rotating
  system that is thermally stable against axisymmetric perturbations
  is stable also against non-axisymmetric perturbations} \citep[see
  also][]{Sun75}.  Similarly, when $\omegad<0$ {\it a uniformly
  rotating system that is thermally unstable against axisymmetric
  perturbations is unstable also against non-axisymmetric
  perturbations}.

In order to illustrate the effect of varying $\kphi^2/k^2$, in
Fig.~\ref{fig:nonax} we plot the results for representative cases with
$\gammaprime<\gamma$ (namely $\gammaprime=1$; solid arrows) and
$\gammaprime>\gamma$ (namely $\gammaprime=3$; long dashed arrows),
assuming $\kz/\kR=2$, $|\omegad|/\Omega=0.3$, $\czerotil=2$,
$\Gammapz=-1.5$, and $\GammapR=-1.25$.  Of course the specific
behaviour of the system depends on the choice of the parameters,
however the examples shown in Fig.~\ref{fig:nonax} represent
qualitatively the trends of typical models in the four quadrants.
When all the other parameters are fixed, the modulus of $\omegaBVsq$
increases for increasing $\kphi^2/k^2$. Therefore, when the system is
convectively stable ($\omegaBVsq>0$) increasing $\kphi$ (i.e.,
considering non-axisymmetric perturbations with larger $m$ or shorter
azimuthal wavelength) moves our system towards the top-left of the
first quadrants of the diagrams in Fig.~\ref{fig:nonax}.  It is
interesting to note that when $|\kphi/\kz|\to \infty$, the system can
become overstable, because $\omegarotsq\to 0$ (we recall that for
$\omegad<0$ the system is overstable if $\omegarotsq=0$ and
$\omegaBVsqtil>\frac{1}{4}$).  Systems with $\omegaBVsq<0$ move, for
increasing $\kphi$, towards the bottom-left part of the diagrams in
Fig.~\ref{fig:nonax}: in these cases, non-axisymmetric effects cannot
change the unstable nature of a system with $\omegad<0$, but it is
interesting to note that a system with $\omegad>0$, overstable against
axisymmetric perturbations, can be unstable against high-$m$
non-axisymmetric modes.

\subsection{Differential rotation}
\label{sec:diff}

Here we consider the case of non-axisymmetric perturbations in a
differentially rotating corona, which is complicated by the effect of
the shear on the perturbations. A possibility would be to consider
disturbances in the form $\Fzero+F(R,z)\exp(-\i\omega t+\i m \phi)$,
where $\Fzero$ is the unperturbed quantity and $|F|\ll|\Fzero|$
\citep[see e.g.][]{Cow51,Lin64,Sun74a,Sun75,Fuj87,Han87,Ber89}:
linearizing the hydrodynamic equations with these perturbations leads
in general to a system of partial differential equations.  As an
alternative, one can adopt shearing coordinates \citep{Gol65} to
reduce the problem to a system of ordinary differential equations
\citep[see also][]{BalH92,Bra06,Bal09}.  Here we follow the latter
approach, and in particular we adopt the formalism used by
\cite{BalH92} in their study of non-axisymmetric perturbations in a
differentially rotating magnetized disc.

Let us consider an unperturbed distribution that is a solution of
equations~(\ref{eq:hydrocylc}-\ref{eq:hydrocyle}) with vanishing time
partial derivatives and $\vvzero=(0,0,\vzerophi)$, where
$\vzerophi=\Omega(R,z)R$ (for simplicity we assume here
$\vzeroR=\vzeroz=0$).  Following \citet{Gol65} we perform the change
of coordinates $\phi'=\phi-\Omega(R,z)t$, $R'=R$, $z'=z$ and $t'=t$.
The partial derivatives are transformed as follows:
\begin{eqnarray}
&&{\pd \over \pd \phi}={\pd \over \pd \phi'},\\
&&{\pd \over \pd R}={\pd \over \pd R'}-t{\pd \Omega \over \pd R}{\pd \over \pd \phi'},\\
&&{\pd \over \pd z}={\pd \over \pd z'}-t{\pd \Omega \over \pd z}{\pd \over \pd \phi'},\\
&&{\pd \over \pd t}={\pd \over \pd t'}-\Omega(R,z){\pd    \over \pd \phi}.
\end{eqnarray}
We perturb the system with small non-axisymmetric disturbances.  In
the primed coordinates the perturbation takes the form of a plane
wave, so we can write the perturbed quantities as
$\Fzero+F(t')\exp(\i\kR' R' + \i\kz' z'+ \i m \phi')$, with $\kR'$,
$\kz'$ and $m$ constant.  It is convenient to work in the Boussinesq
approximation: linearizing
equations~(\ref{eq:hydrocylc}-\ref{eq:hydrocyle}) we get
\begin{eqnarray}
 &&\kR(t')\vR+\kz(t')\vz+\kphi\vphi =0,\label{eq:nonaxdiffc}\\
 &&{\d \vR\over \d t'}\rhozero+\i\kR(t') p-\ApR\czero^2\rho-2\Omega\vphi\rhozero=0,\\
 && {\d \vz\over \d t'}\rhozero+\i\kz(t')p-\Apz\czero^2\rho=0,\\
 && {\d\vphi\over \d t'}\rhozero+\vR\rhozero(\Omega+\OmegaR)+\vz\rhozero\Omegaz+\i \kphi p=0,\\
&&-\gamma{\d\rho\over \d t'}{1\over\rhozero}
+\vR\left(\ApR-\gamma\ArhoR\right)+\vz\left(\Apz-\gamma\Arhoz\right)-\gamma{\rho\over \rhozero}[\omegac(t')+\omegath]=0,
\label{eq:nonaxdiffe}
\end{eqnarray}
where $\kphi\equiv m/R'$,
\begin{equation}
\kR(t')=\kR'-m t' {\partial \Omega \over \partial R}
\end{equation} 
and
\begin{equation}
\kz(t')=\kz'-m t' {\partial \Omega \over \partial z}.
\end{equation}
Note that now $\omegac$, still defined as in
equation~(\ref{eq:omegac}), depends on time, because it depends on
$k^2(t')=\kR(t')^2+\kz(t')^2+\kphi^2$.  As implementations of the
Boussinesq approximation we neglected the term proportional to ${\d
  \rho/ \d t'}$ in equation~(\ref{eq:nonaxdiffc}) and the term
proportional to ${\d p/ \d t'}$ in equation~(\ref{eq:nonaxdiffe}), and
we assumed $\rhozero T= -\Tzero\rho$.

The system of five ordinary differential
equations~(\ref{eq:nonaxdiffc}-\ref{eq:nonaxdiffe}) in the unknown
$\vR$, $\vz$, $\vphi$, $\rho$ and $p$ can be simplified by eliminating
$\vphi$ and $p$. In particular, in the momentum equations, $\vphi$ can
be eliminated by using the mass conservation equation (and its time
derivative):
\begin{equation}
\vphi=-{\kR(t') \over \kphi}\vR-{\kz(t') \over \kphi}\vz, 
\end{equation}
\begin{equation}
{\d \vphi \over \d t'}=-{\kR(t')\over \kphi}{\d \vR \over \d t'}-{\kz(t')\over \kphi}{\d \vz \over \d t'}.
\end{equation}
The pressure perturbation $p$ can be eliminated by combining the
momentum equations as follows: we subtract the $z$ equation
(multiplied by $\kR$) from the $R$ equation (multiplied by $\kz$), and
we subtract the $z$ equation (multiplied by $\kphi$) from the $\phi$
equation (multiplied by $\kz$).  After rearrangement and
simplification, we end up with the following system of three coupled
ordinary differential equations (in the variables $\vR$, $\vz$ and
$\rho$), which fully describes the evolution of the non-axisymmetric
perturbations in a differentially rotating corona:
\begin{eqnarray}
&&{\d \vR \over \d t}=
{2 \kR \kphi \Omega \over k^2}\left({\partial \ln \Omega \over \partial \ln R }-{\kz^2\over\kphi^2}\right)\vR+
{2 \kR \kphi \Omega \over k^2}\left({R\over z}{\partial \ln \Omega \over \partial \ln |z| }-{\kz\over\kR}{\kz^2+\kphi^2 \over \kphi^2}\right)\vz+
{\czero^2 \ApR\over \rhozero}\left({\kz^2+\kphi^2\over k^2}-{\Apz \over\ApR }{\kR\kz\over k^2}\right)\rho,\\
&&{\d \vz \over \d t}=
{2 \kz \kphi \Omega \over k^2}\left({\partial \ln \Omega \over \partial \ln R }+{\kR^2+\kphi^2\over\kphi^2}\right)\vR+
{2 \kz \kphi \Omega \over k^2}
\left(
{R\over z}
{\partial \ln \Omega \over \partial \ln |z| }+
{\kz\kR\over\kphi^2}\right)\vz+
{\czero^2 \Apz\over \rhozero}\left({\kR^2+\kphi^2\over k^2}-{\ApR \over\Apz }{\kR\kz\over k^2}\right)\rho,\\
&&{\d\rho\over \d t}=
{\rhozero\over \gamma}\left(\ApR-\gamma\ArhoR\right)\vR+
{\rhozero\over \gamma}\left(\Apz-\gamma\Arhoz\right)\vz-
\left(\omegac+\omegath \right)\rho,
\label{eq:system}
\end{eqnarray}
where we dropped the explicit time dependence of $\kR$, $\kz$ and
$\omegac$.  It is useful to rewrite the system above more compactly
using only dimensionless quantities:
\begin{eqnarray}
&&{\d \vRtil \over \d \tau}=
{2 \kRtil \over \ktil^2}\left(\GammaOmegaR-\kztil^2\right)\vRtil+
{2 \kRtil \over \ktil^2}\left[\GammaOmegaz-{\kztil\over\kRtil}(\kztil^2+1)\right]\vztil+
{\czerotil^2 \GammapR}\left({\kztil^2+1\over \ktil^2}-{\Gammapz \over\GammapR }{\kRtil\kztil\over \ktil^2}\right)\rhotil,\label{eq:adimvr}\\
&&{\d \vztil \over \d \tau}=
{2 \kztil \over \ktil^2}\left(\GammaOmegaR+\kRtil^2+1\right)\vRtil+
{2 \kztil \over \ktil^2}
\left(
\GammaOmegaz+
\kztil\kRtil\right)\vztil+
{\czerotil^2 \Gammapz}\left({\kRtil^2+1\over \ktil^2}-{\GammapR \over\Gammapz }{\kRtil\kztil\over \ktil^2}\right)\rhotil,\label{eq:adimvz}\\
&&{\d\rhotil\over \d \tau}=
{{1 \over \gamma}\left(\GammapR-\gamma\GammarhoR\right)}\vRtil+
{{1 \over \gamma}\left(\Gammapz-\gamma\Gammarhoz\right)}\vztil-
(\omegactil \ktil^2+\omegathtil)\rhotil,
\label{eq:adimrho}
\end{eqnarray}
where $\vRtil\equiv\vR/\Omega R$, $\vztil\equiv\vz/\Omega R$,
$\rhotil\equiv\rho/\rhozero$, $\tau\equiv t' \Omega$, $\kRtil\equiv
\kR/\kphi$, $\kztil\equiv \kz/\kphi$, $\ktil\equiv k/\kphi$,
\begin{eqnarray}
 &&\GammarhoR\equiv{\pd \ln \rhozero \over \pd \ln
    R}=R\ArhoR,\qquad \Gammarhoz\equiv{R\over z}{\pd \ln \rhozero
    \over \pd \ln |z|}=R\Arhoz,\\
&&\GammaOmegaR\equiv{\pd \ln \Omega \over \pd \ln R},\qquad \GammaOmegaz\equiv{R\over z}{\pd \ln \Omega \over \pd \ln |z|},
\end{eqnarray}
$\GammapR$ and $\Gammapz$ are defined in equations~(\ref{eq:Gammap}),
\begin{equation}
\omegactil\equiv{\omegac \kphi^2\over \Omega k^2}= {1\over
  \Omega}\left({\gamma-1\over\gamma}\right){\kphi^2 \kappa
  \Tzero^{7/2}\over \pzero},\label{eq:omegactil}
\end{equation}
and
\begin{equation}
\omegathtil\equiv{\omegath \over \Omega}\label{eq:omegathtil}.
\end{equation}
In equations~(\ref{eq:adimvr}-\ref{eq:adimrho}), besides the unknown
$\vRtil(\tau)$, $\vztil(\tau)$ and $\rhotil(\tau)$, the only other
time-dependent quantities are
\begin{equation}
\kRtil(\tau)=\kRtil'-\GammaOmegaR\tau,
\end{equation}
\begin{equation}
 \kztil(\tau)=\kztil'-\GammaOmegaz\tau,
\end{equation}
and $\ktil^2(\tau)=\kRtil^2+\kztil^2+1$, where $\kRtil'\equiv
\kR'/\kphi$ and $\kztil'\equiv \kz'/\kphi$ (note that $\omegactil$
does not depend on time). The system of ordinary differential
equations is thus completed by specifying at $\tau=0$ the values of
$\vRtil$, $\vztil$ and $\rhotil$, and by choosing the values of the
following parameters: $\kRtil'$, $\kztil'$, $\GammaOmegaR$,
$\GammaOmegaz$, $\GammapR$, $\Gammapz$, $\GammarhoR$, $\Gammarhoz$,
$\czerotil$, $\omegactil$ and $\omegathtil$. A full exploration of the
parameter space is prohibitive, so we will present solutions for
specific relevant cases, obtained by numerically integrating the
system of equations~(\ref{eq:adimvr}-\ref{eq:adimrho}) with a
fourth-order Runge-Kutta method.

\subsubsection{Barotropic distributions}
\label{sec:barotr}

\begin{figure}
\centerline{ \psfig{file=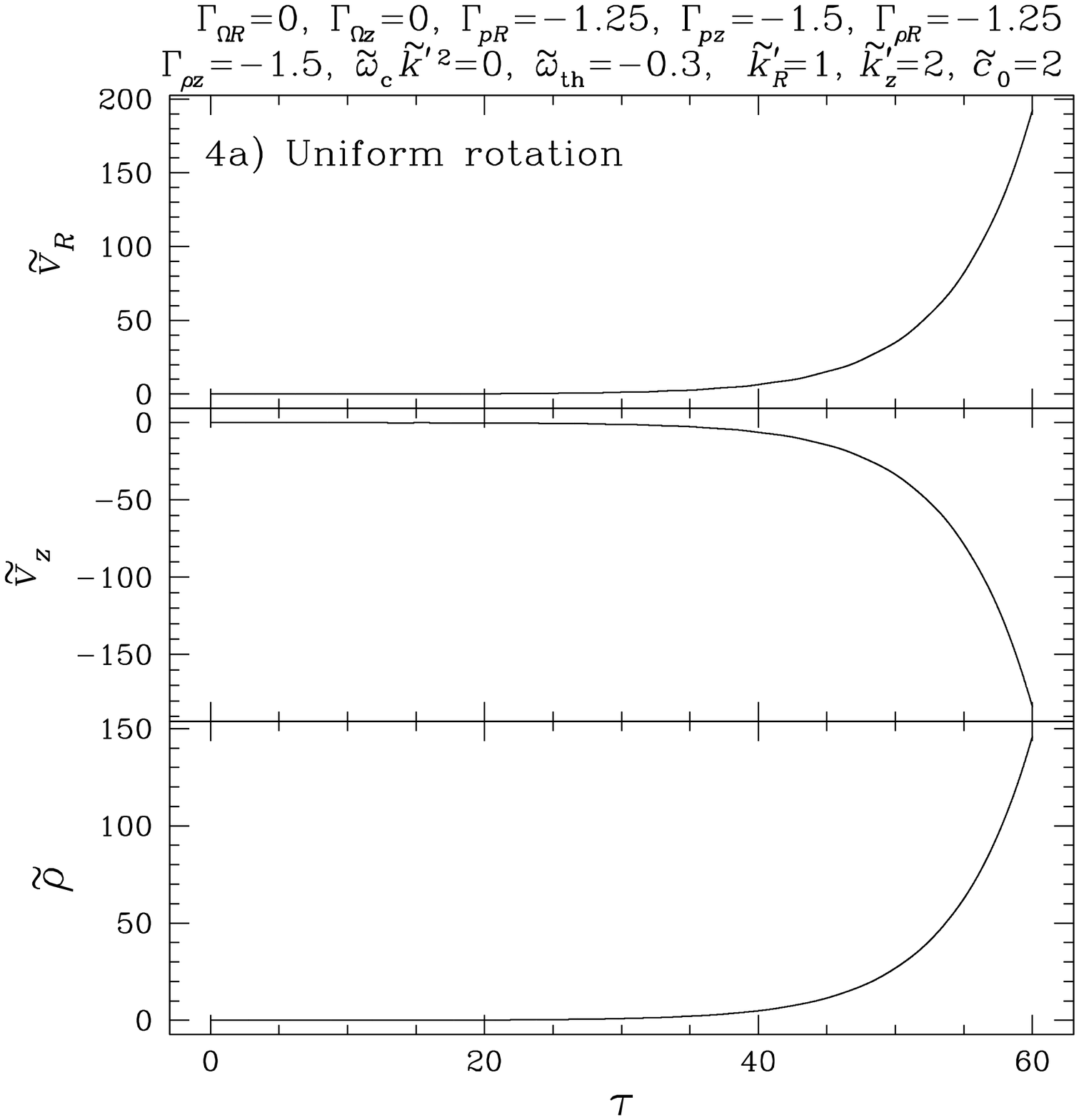,width=0.5\hsize}
  \psfig{file=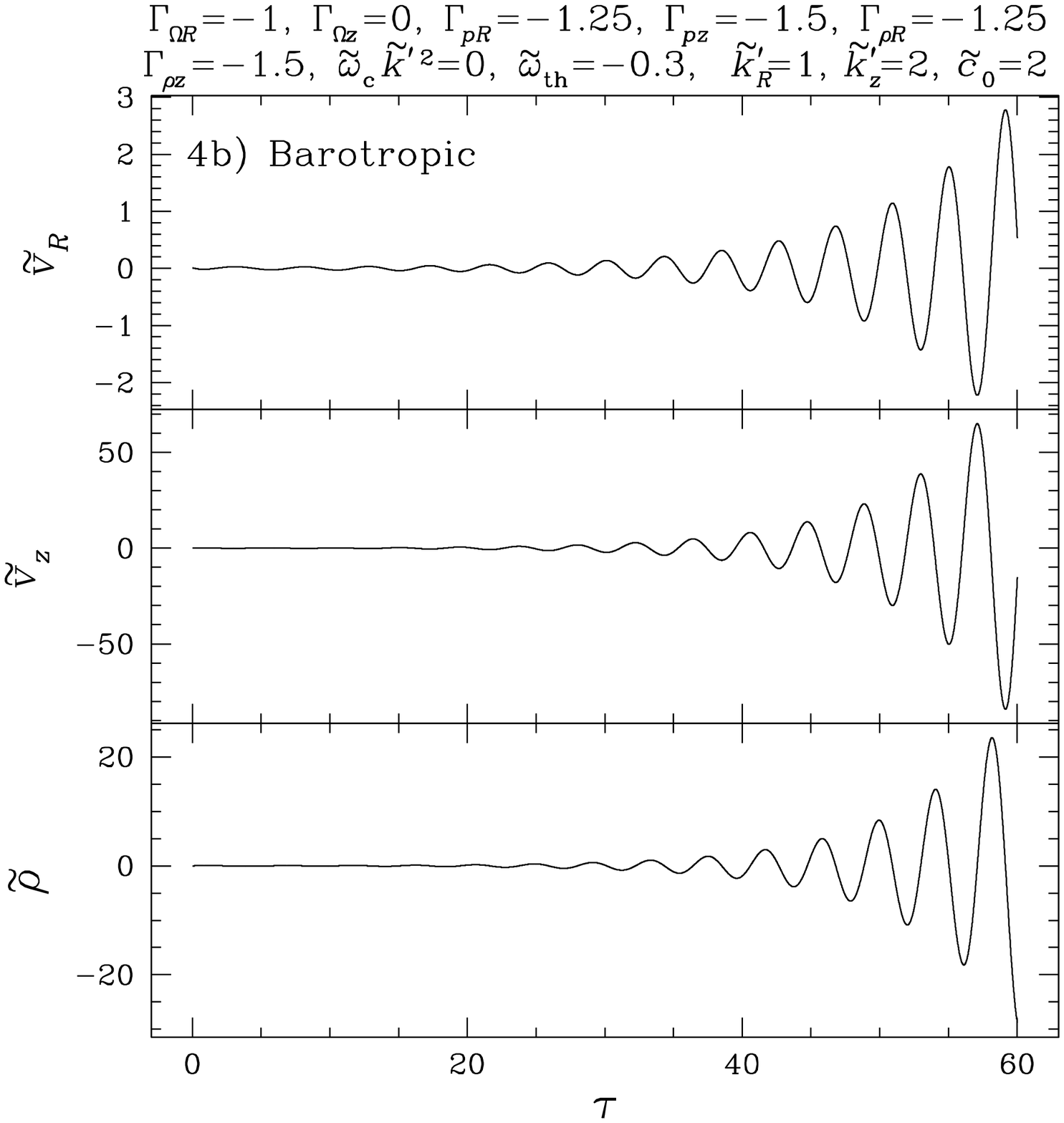,width=0.5\hsize} } \centerline{
  \psfig{file=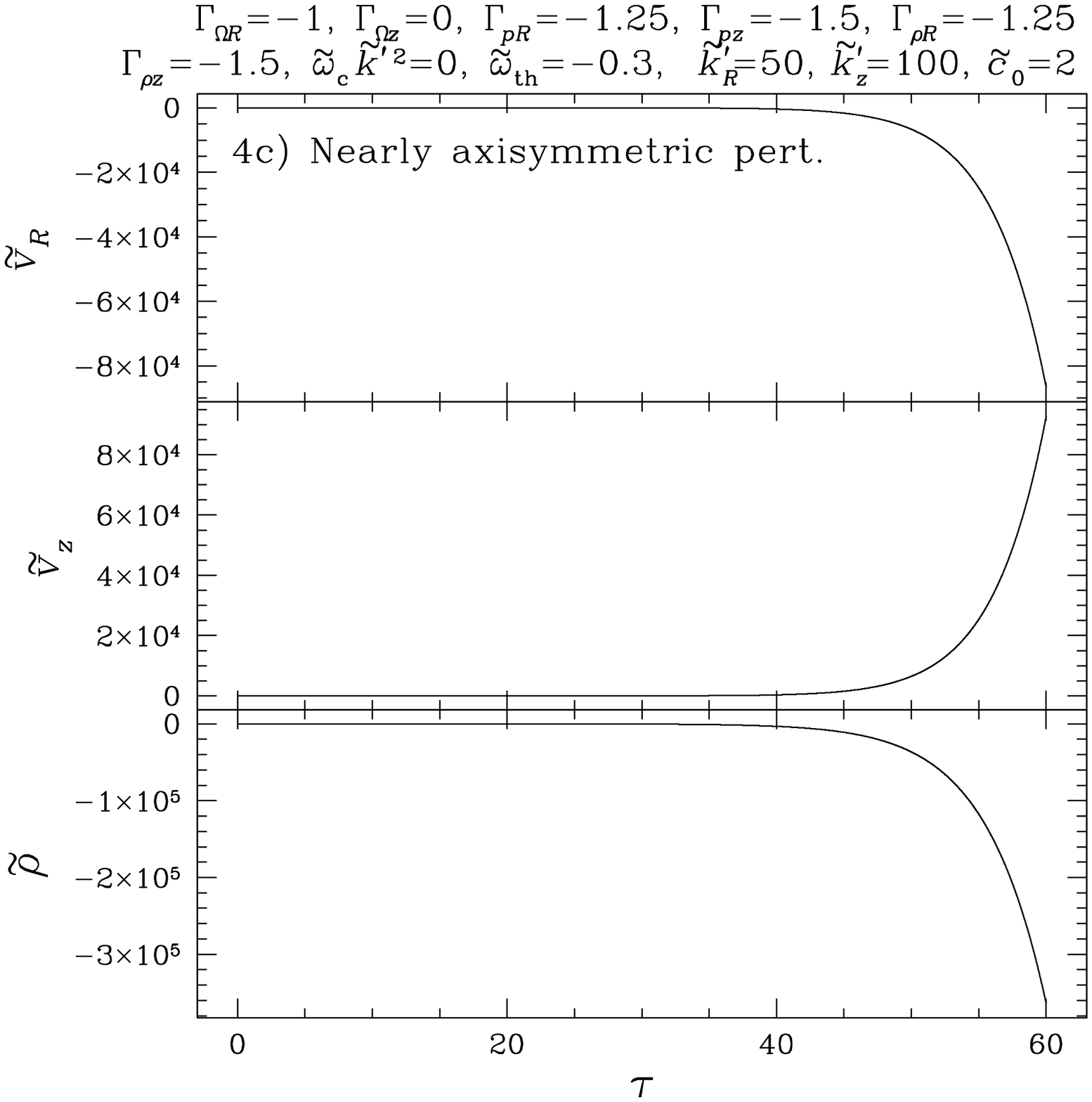,width=0.5\hsize}
  \psfig{file=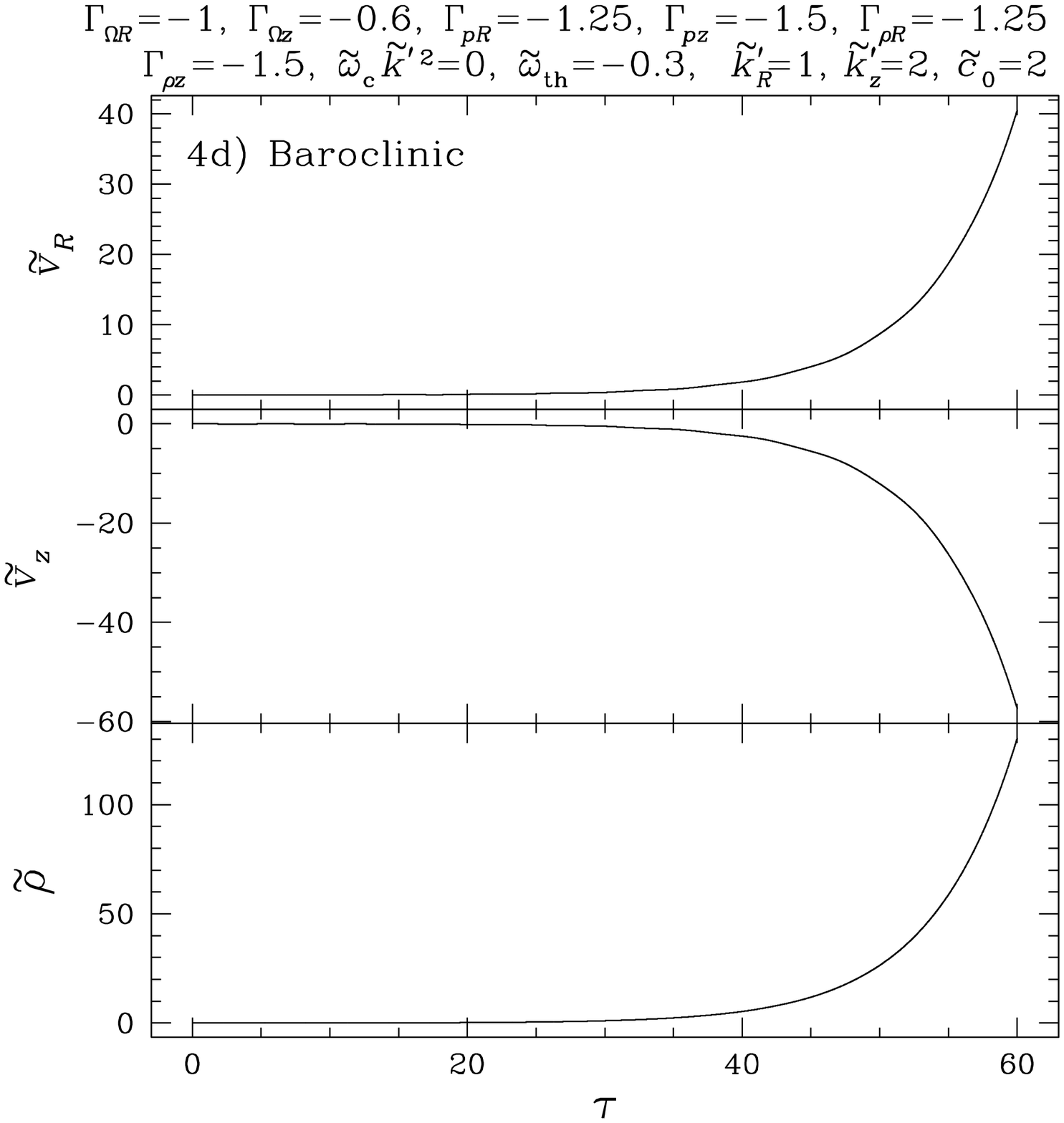,width=0.5\hsize} }
\caption{Time evolution of non-axisymmetric velocity and density
  perturbations for rotating stratified fluids in the presence of
  radiative cooling ($\omegathtil=-0.3$), but in the absence of
  thermal conduction ($\omegactil=0$). In all cases we assume
  $\vRtil(0)=0.01$, $\vztil(0)=0.01$, $\rhotil(0)=0.01$,
  $\kztil'/\kRtil'=2$, $\czerotil=2$, $\Gammapz=-1.5$, and
  $\GammapR=-1.25$. In model 4a the system is uniformly rotating
  ($\GammaOmegaR=0$, $\GammaOmegaz=0$), and the perturbation has
  ``low'' azimuthal wave-number ($\kRtil'=1$).  Model 4b is the same
  as model 4a, but with $\GammaOmegaR=-1$ (barotropic, differentially
  rotating, ``low'' azimuthal wave-number). Model 4c is the same as
  model 4b, but with $\kRtil'=50$ (barotropic, differentially
  rotating, ``high'' azimuthal wave-number). Model 4d is the same as
  model 4b, but with $\GammaOmegaz=-0.6$ (baroclinic, differentially
  rotating, ``low'' azimuthal wave-number). Time is in units of
  $\Omega^{-1}$.}
\label{fig:odeomegath}
\end{figure}

Let us focus first on solutions of the system of
equations~(\ref{eq:adimvr}-\ref{eq:adimrho}) in the case of barotropic
distributions (see Section~\ref{sec:eq}), for which $\Omega=\Omega(R)$
and $\pzero=\pzero(\rhozero)$, with local polytropic index $
\gammaprime=\d \ln \pzero/ \d \ln \rhozero$.  We are interested in
estimating the effect of the non-axisymmetry of the perturbations and
of the differential rotation on the thermal stability of the fluid, so
it is convenient to make a comparison with the results obtained for
axisymmetric perturbations (Section~\ref{sec:axisymm}) and uniform
rotation (Section~\ref{sec:unif}). In those cases, we found it useful
to explore separately the cases $\omegad=0$, $\omegad>0$ and
$\omegad<0$, where $\omegad=\omegath+\omegac$ is the characteristic
frequency of dissipative processes. In the case of non-axisymmetric
perturbations and differential rotation $\omegad=\Omega(\omegactil
\ktil^2+\omegathtil)$ depends on time through $\ktil(t')$, so for
comparison with the previous analysis it is convenient to distinguish
the following cases:
\begin{enumerate}
\item {\it Case with $\omegactil=0$ and $\omegathtil=0$}.  This is the
  non-dissipative case, which has been studied in previous work on
  convective instability in rotating stratified fluids
  \citep{Cow51,Sun75}.  In accordance with \cite{Sun75}, we find that
  the combination of differential rotation and non-axisymmetry of the
  perturbations can have a destabilizing effect against
  convection. For instance, we find that---for sufficiently large
  values of $\kphi$---adiabatic ($\gammaprime=\gamma$) configurations
  are stable for $\GammaOmegaR=0$ (uniform rotation), but are unstable
  for $\GammaOmegaR\neq 0$ (differential rotation), even when the
  specific angular momentum increases outwards ($\GammaOmegaR>-2$).

\item {\it Case with $\omegactil>0$ and $\omegathtil=0$}. This is a
  dissipative case, with positive dissipative term, analogous to the
  model with radiative diffusion studied by \citet{Sun75}.  For
  axisymmetric perturbations or uniform rotation, when $\omegad>0$ the
  only stable configurations are those characterized by
  $\gammaprime<\gamma$ and $d(\Omega R^2)/d R>0$
  (i.e. $\GammaOmegaR>-2$); the other configurations are unstable or
  overstable (see left-hand panels of Figs.~\ref{fig:axisymm} and
  \ref{fig:nonax}). We calculated numerical solutions for
  configurations with $\gammaprime<\gamma$ and $\GammaOmegaR>-2$
  (known to be stable against axisymmetric perturbations), finding
  them stable also against non-axisymmetric perturbations for a wide
  range of values of $\kRtil$ and $\kztil$ \citep[see
  also][]{Sun75}. 

\item {\it Case with $\omegactil=0$ and $\omegathtil<0$}.  This is a
  dissipative case, with negative dissipative term, to be compared
  with the cases of axisymmetric perturbations or uniform rotation
  with $\omegad<0$ (right-hand panels of Figs.~\ref{fig:axisymm} and
  \ref{fig:nonax}).  In those cases, the systems were found unstable
  in the relevant parameter regime $\gammaprime<\gamma$ and
  $\GammaOmegaR>-2$ (outward-increasing specific entropy and angular
  momentum). We explored the stability against non-axisymmetric
  perturbations for several configurations with $\gammaprime<\gamma$
  and $\GammaOmegaR>-2$ finding that {\it the combination of
    non-axisymmetry of the perturbations and differential rotation can
    turn instability into overstability}. This can be seen from
  Fig.~\ref{fig:odeomegath}, showing, for example, the time evolution
  of non-axisymmetric perturbations in three configurations (models
  4a, 4b and 4c) differing only for the values of $\GammaOmegaR$
  (measuring the degree of differential rotation), $\kRtil'$ and
  $\kztil'$ (measuring the deviation from axisymmetry of the
  perturbations). The modes considered in models 4a and 4b have
  $\kphi$ of the order of $\kR$ and $\kz$, so they are relevant to
  blob-like perturbations.  In model 4a rotation is uniform, and the
  system is unstable, consistent with the calculations of
  Section~\ref{sec:unif}. Model 4b differs from model 4a only because
  differential rotation (with $\GammaOmegaR=-1$) replaces uniform
  rotation: as a consequence the system is overstable. It must be
  noted that the growth rate of the amplitude of the overstable mode
  shown in panel 4b is not negligible: the perturbation, which is
  assumed of the order of $1\%$ at $\tau=0$, becomes nonlinear at
  $\tau \sim 25$.  Model 4c also rotates differentially with
  $\GammaOmegaR=-1$, as model 4b, but now the value of the azimuthal
  wave-number is smaller by a factor of 50 than in model 4b, meaning
  that the perturbation is nearly axisymmetric: this mode, which is
  relevant for disturbances very elongated along $\phi$, is clearly
  unstable.

\item {\it Case with $\omegactil>0$ and $\omegathtil<0$}.  In this
  case both thermal conduction and cooling are present. Let us focus
  again on cases in which $\gammaprime<\gamma$ and $\GammaOmegaR>-2$
  (outward-increasing specific entropy and angular
  momentum). Considering the initial value of the dissipative
  frequency $\omegad(0)=\Omega(\omegactil\ktil'^2+\omegathtil)$, where
  $\ktil'^2\equiv\kRtil'^2+\kztil'^2+1=\ktil^2(0)$, we distinguish
  different cases according to the sign of $\omegad(0)$.  When
  $\omegad(0)>0$, all explored cases of non-axisymmetric perturbations
  in the presence of differential rotation turn out to be stable, so
  differential rotation and deviation from axisymmetry do not have
  destabilizing effects on configurations stabilized by thermal
  conduction. When $\omegad(0)<0$ we find stability, overstability or
  instability depending on the values of the parameters. Given that in
  the limit of uniform rotation or axisymmetric perturbations we
  always have instability in this parameter regime (see right-hand
  panels of Figs.~\ref{fig:axisymm} and \ref{fig:nonax}), it is clear
  that {\it non-axisymmetry of the perturbations and differential
    rotation can have a stabilizing effect}. An example is shown in
  Fig.~\ref{fig:odeomegac}, plotting the time evolution of
  non-axisymmetric perturbations in three configurations (5a, 5b and
  5c), having the same values parameters as the corresponding
  configurations considered in Fig.~\ref{fig:odeomegath}, but
  $\omegactil k'^2=0.1$, so that the initial value of the dissipative
  frequency is $\omegad(0)=-0.2\Omega$. The uniformly rotating case
  (model 5a) is unstable, consistent with the results of
  Section~\ref{sec:unif}.  Interestingly, the differentially rotating
  case with the same wave-vector (model 5b) is stable: in this case
  $\omegad(t)$ is negative at $t=0$, but soon becomes positive because
  $\kRtil^2$ increases as a consequence of differential rotation
  (making also $\ktil^2$ and $\omegac$ increase). We recall that the
  analogous configuration, in the absence of conduction is overstable
  (model 4b in Fig.~\ref{fig:odeomegath}).  Physically, the shear
  tends to distort any overdense region by making it narrow in the $R$
  direction, so the growth of the perturbation is more easily damped
  by thermal conduction. As models 4a and 4b, also models 5a and 5b
  are characterized by modes with $\kphi$ of the order of $\kR$ and
  $\kz$, so they are of interest for blob-like perturbations. If the
  azimuthal wave-number is small enough (i.e. the mode is almost
  axisymmetric, relevant to overdensities very elongated along $\phi$)
  the system appears unstable even in the presence of differential
  rotation (model 5c). Formally, also in this case $\omegad(t)$
  becomes positive over sufficiently long times, and eventually the
  perturbation is damped. However, this must not necessarily happen in
  reality, because it is clear that the linear analysis breaks when at
  least one among $\vRtil$, $\vztil$ and $\rhotil$ becomes much larger
  than unity.
\end{enumerate}

\subsubsection{Baroclinic distributions}
\label{sec:barocl}

In the previous Section we focused on barotropic distributions,
i.e. configurations in which $\Omega=\Omega(R)$ and
$\pzero=\pzero(\rhozero)$.  We consider here the problem of the
stability against non-axisymmetric perturbations of systems with
baroclinic distributions: in other words, we allow for the fact that
$\Omega$ can depend on $z$ as well as on $R$ (implying that $\pzero$
is not stratified with $\rhozero$; see Section~\ref{sec:eq}). In terms
of the formalism here adopted, we describe solutions of
equations~(\ref{eq:adimvr}-\ref{eq:adimrho}) in cases with
$\GammaOmegaz\neq0$.

As it happens for the barotropic cases, also baroclinic configurations
are found to be stable, unstable or overstable depending on the
specific choice of the parameters.  In general we find that adding a
relatively small vertical gradient of $\Omega$ does not change the
stability properties of a system.  However, we found that in some
cases the fact that $\Omega$ depends on $z$ can have destabilizing
effects. In order to isolate this phenomenon, we considered models
with the same parameters as the barotropic models discussed above, but
with $\GammaOmegaz\neq0$.  For small values of $|\GammaOmegaz|$ the
baroclinic models behave like the corresponding barotropic models,
but---in the absence of thermal conduction---instability replaces
overstability for sufficiently large $|\GammaOmegaz|$
(i.e. sufficiently strong dependence on $z$ of the angular velocity).
An example of thermally unstable baroclinic configuration is model 4d
in Fig.~\ref{fig:odeomegath}, which has the same values of the
parameters as the (overstable) barotropic model 4b, but
$\GammaOmegaz=-0.6$.  When thermal conduction is present baroclinic
models are found to be stable if the ``corresponding'' barotropic
models are stable: compare, in Fig.~\ref{fig:odeomegac}, the
barotropic model 5b with the baroclinic model 5d.

\section{Implications for galactic coronae}
\label{sec:impl}

Here we use the results obtained above to address the question of
whether the coronae of disc galaxies can fragment, via thermal
instability, into cool, pressure-supported clouds, which, in the case
of the Milky Way, would be identified with the high-velocity clouds.
As already pointed out, the physical properties of the galactic
coronae of disc galaxies are poorly constrained observationally. Such
coronae are expected to be stratified, almost in equilibrium at the
system's virial temperature, similar to the hot atmospheres of massive
elliptical galaxies and galaxy clusters, but characterized by
significantly lower gas temperature and density, and possibly by
non-negligible rotation.  

MRB studied the problem of the thermal stability of the hot
atmospheres of galaxy clusters, using Eulerian plane-wave
perturbations and assuming that the gas does not rotate, and concluded
that these systems are not prone to significant thermal instability.
It has been pointed out that when a background flow is present, the
use of Lagrangian perturbations is preferable to that of Eulerian
perturbations \citep{Bal88}.  However, the main results of MRB have
been confirmed by the Lagrangian study of \cite{BS89}. BNF applied the
thermal-stability analysis of MRB to non-rotating models of the
coronae of disc galaxies, finding that the result obtained for the
highest-temperature atmospheres of clusters (thermal stability or
overstability) apply also to these lower-temperature systems, unless
the gas distribution is perfectly adiabatic (which is unexpected).  In
the present paper we tried to verify whether the conclusions drawn
from the analysis of the non-rotating cases extend also to rotating
coronae. As already stressed, for simplicity we treated the gas as
unmagnetized, so our results apply to real systems only to the extent
that their magnetic fields do not influence substantially the
behaviour of thermal perturbations.

From the calculations reported in the above Sections it is apparent
that rotation introduces some mathematical complexity in the stability
analysis. As a consequence, it is hard to draw general conclusions
about the thermal-stability properties of rotating coronae, which can
be thermally stable, overstable or unstable, depending on the
distribution of specific entropy and angular momentum, but also on the
nature of the perturbations.  If we knew the density, temperature and
specific angular momentum distributions of the corona of the Milky Way
(or of an external disc galaxy), the dispersion relations and the
differential equations derived above could be used to determine
unambiguously whether a given perturbation grows.  Unfortunately, the
uncertainties on the physical properties of the coronae are such that
we cannot be conclusive about their thermal stability or
instability. Nevertheless, some general considerations can be done.

First of all, our calculations show that the stabilizing effect of
thermal conduction found in non-rotating systems occurs also in the
presence of rotation (see cases with $\omegad>0$ in
Sections~\ref{sec:axisymm} and~\ref{sec:nonaxisymm}). As in the
non-rotating case, {\it thermal perturbations with sufficiently short
  wavelength do not grow}. The interesting question is then whether
perturbations with sufficiently long wavelength (i.e. small enough
wave-number $k$, so that $\omegad<0$) grow. It seems reasonable that
galactic coronae are characterized by outward-increasing specific
entropy\footnote{BNF showed that, at least in the absence of rotation,
  very special conditions must be met in order to have a corona with
  flat specific entropy profile.}  and angular momentum, so in the
present discussion we can specialize to configurations with
$\omegaBVsq>0$ and $\omegarotsq>0$.  In Section~\ref{sec:axisymm} we
showed that when $\omegad<0$, such configurations are {\it unstable}
against axisymmetric perturbations, in contrast with analogous
non-rotating configurations, which have been shown to be
overstable. Taken at face value this result indicates that the
presence of even slow rotation can modify qualitatively the stability
properties of a corona: formally, we should conclude that rotating
galactic coronae are thermally unstable, because there is at least one
mode that grows (in fact, all the axisymmetric modes grow).  However,
such instability must not necessarily have important consequences in
the practical problem of the thermal stability of galactic coronae.
We note that the high-velocity clouds, which would be the end-products
of the thermal instability of the Milky Way corona, do not appear
axisymmetric with respect to the rotation axis of the
Galaxy. Moreover, axisymmetric structures of cold gas far from the
plane are not observed, in general, in external disc galaxies.

\begin{figure}
\centerline{
\psfig{file=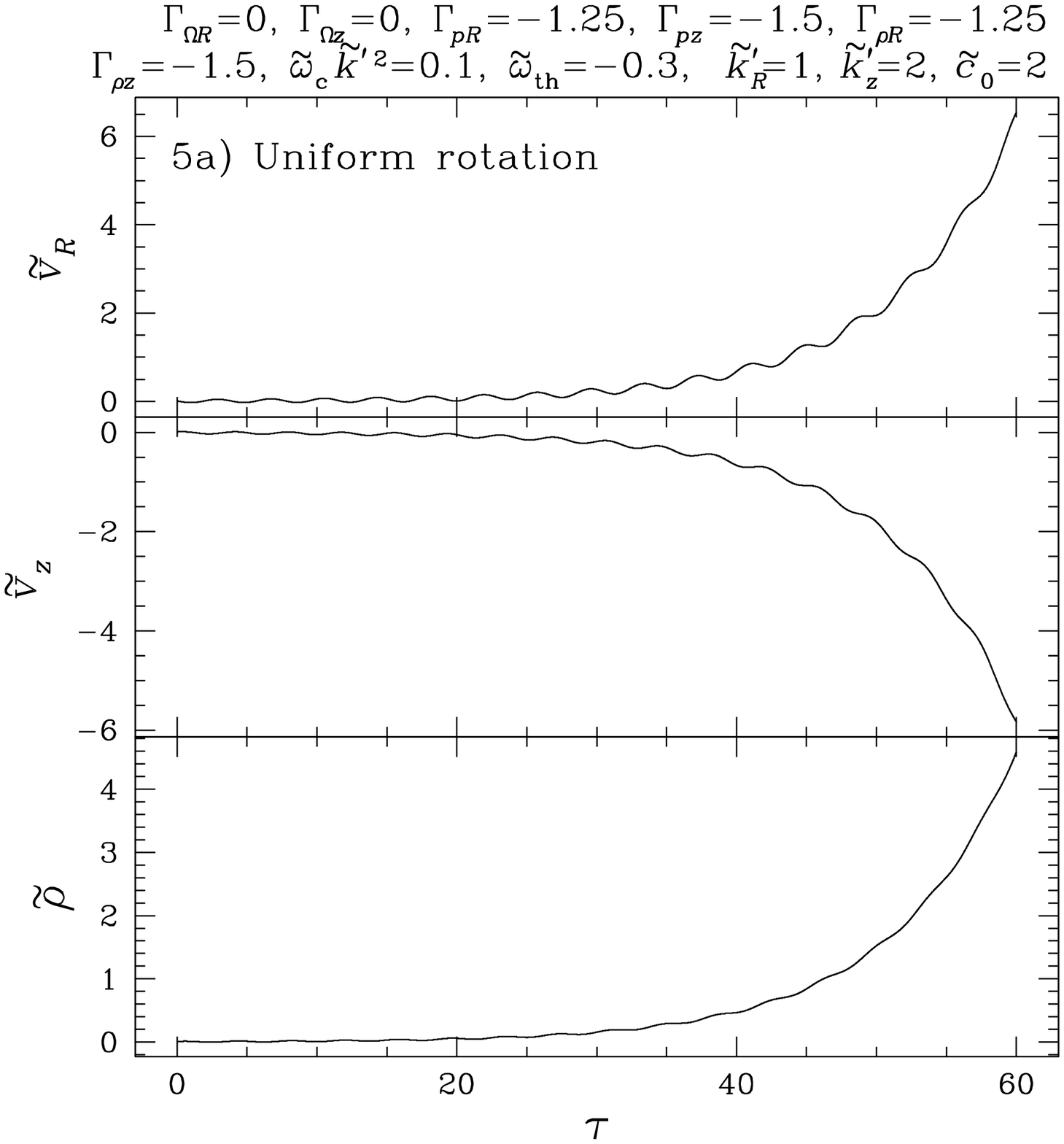,width=0.5\hsize}
\psfig{file=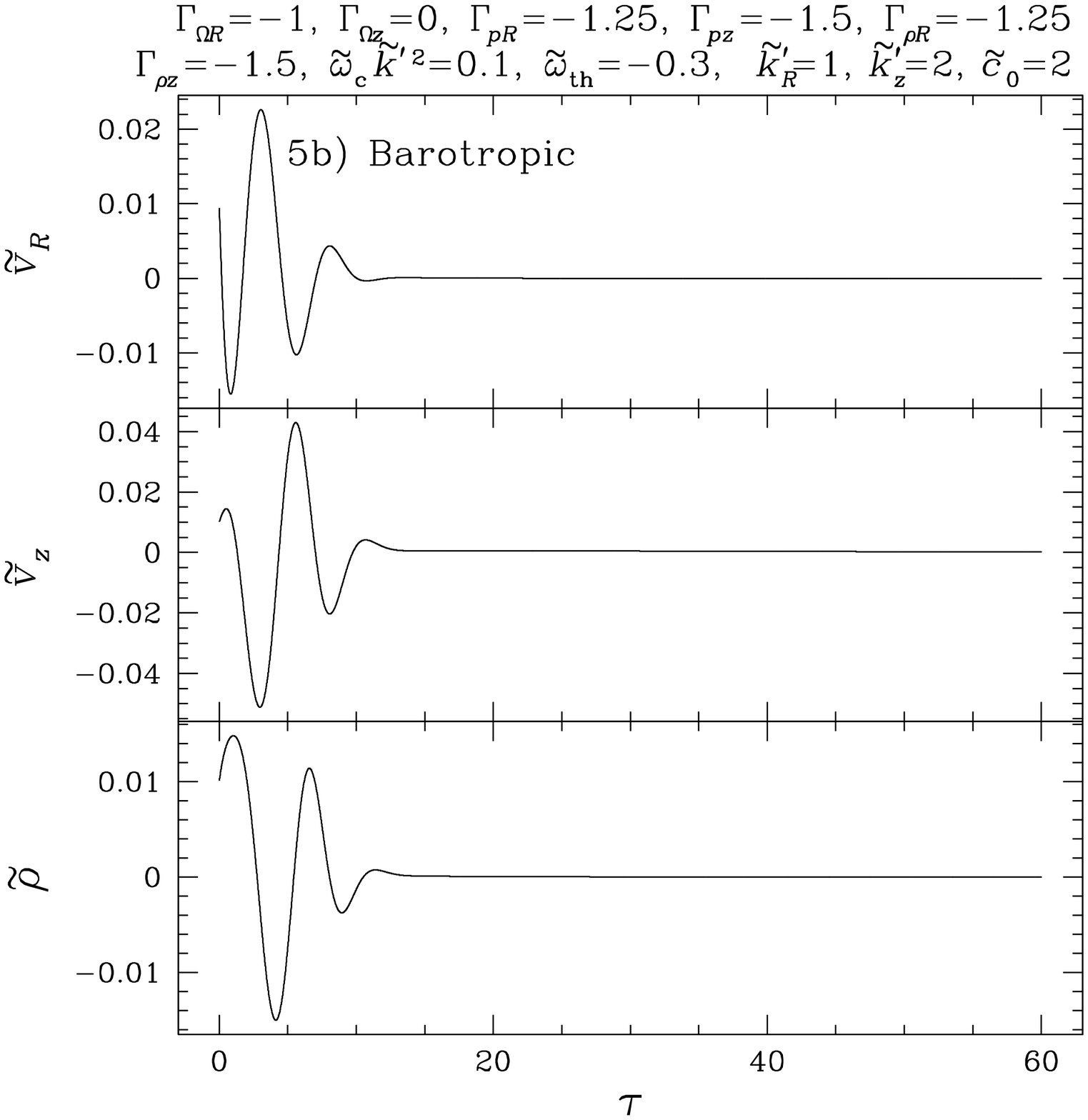,width=0.5\hsize}
}
\centerline{
\psfig{file=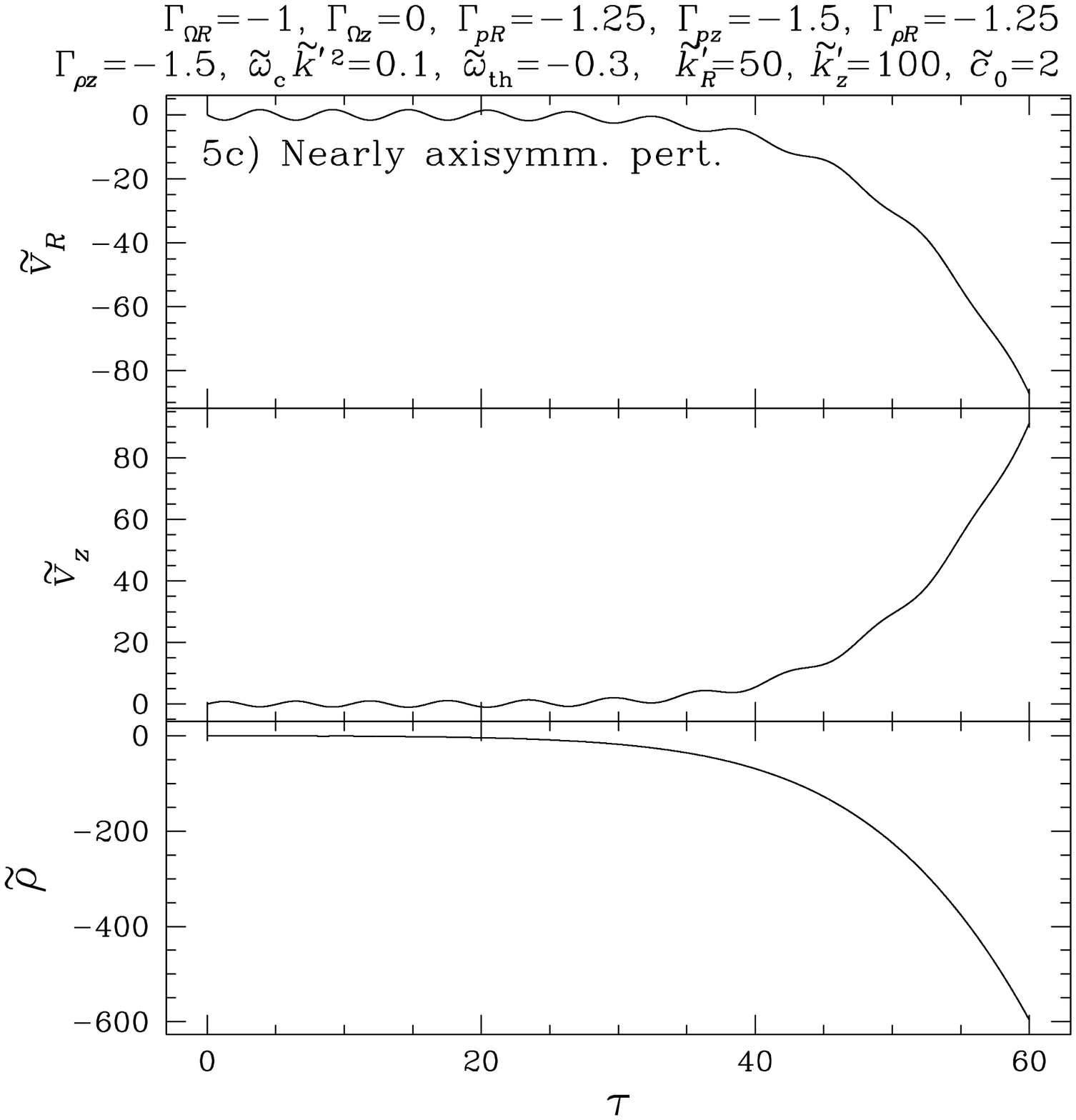,width=0.5\hsize}
\psfig{file=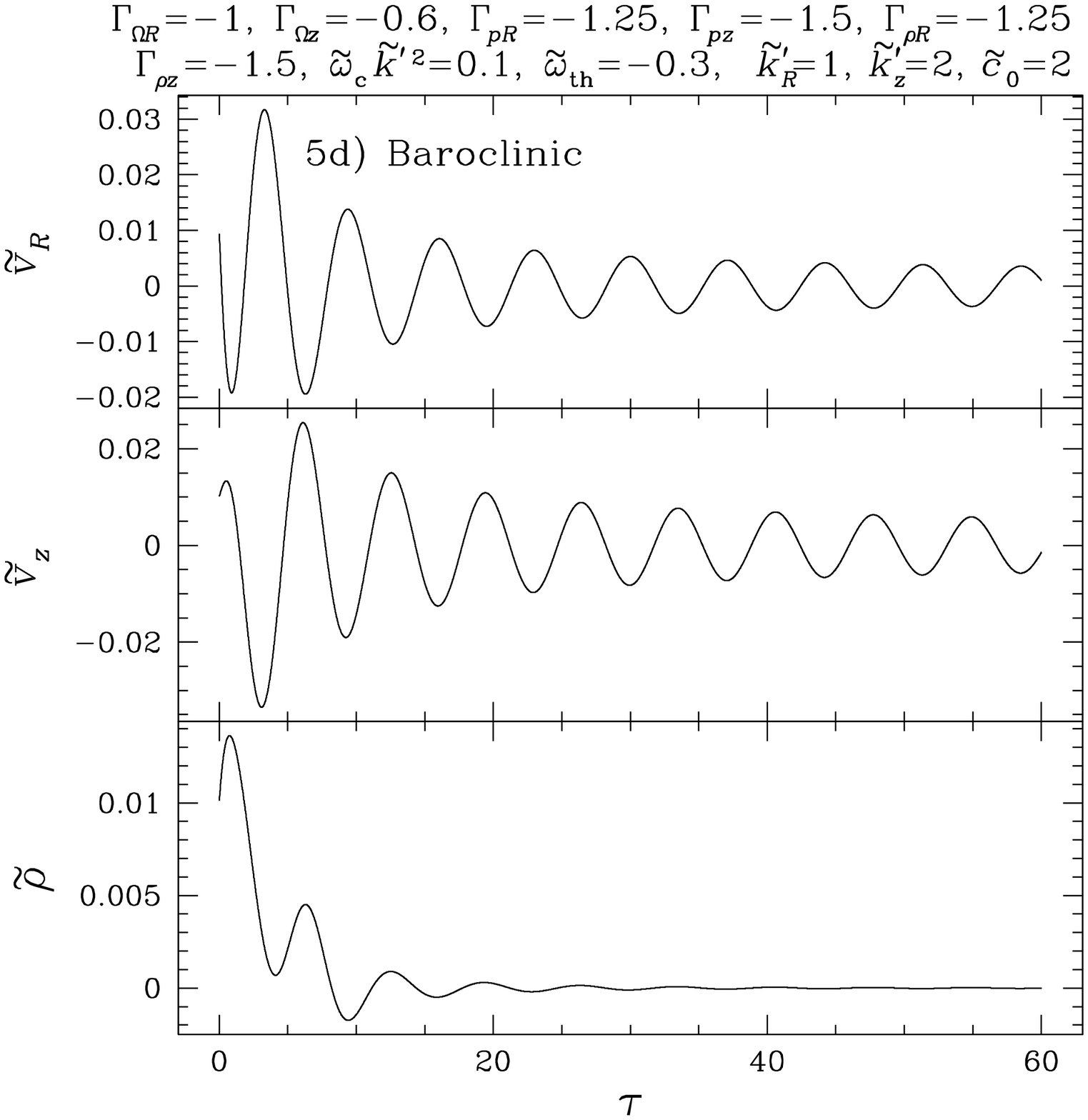,width=0.5\hsize}
}
\caption{Same as Fig.~\ref{fig:odeomegath} for models with all the
  same parameters as the corresponding models in that figure, but with
  $\omegactil>0$. In particular, we adopted $\omegactil k'^2=0.1$,
  where $\ktil'^2\equiv\kRtil'^2+\kztil'^2+1=\ktil^2(0)$, so that the
  initial dissipative frequency $\omegad(0)$ is negative.}
\label{fig:odeomegac}
\end{figure}

It is then interesting to study the behaviour of non-axisymmetric
perturbations, which are relevant to the problem of whether blob-like,
cool clouds can condense out of the corona. We have shown in
Section~\ref{sec:nonaxisymm} that at least some differentially
rotating configurations, which are unstable against axisymmetric
perturbations, are overstable (model 4b in Fig.~\ref{fig:odeomegath})
or stable (model 5b in Fig.~\ref{fig:odeomegac}) against
non-axisymmetric perturbations with sufficiently high azimuthal
wave-number (specifically, to modes with $\kphi$ of the order of $\kR$
and $\kz$, which describe blob-like disturbances). It follows that an
overdense, cool blob in a differentially rotating corona is unlikely
to condense, even if the thermal-instability frequency is faster than
conductive damping (i.e., $\omegad<0$).

From our analysis it also emerged that the distribution of specif
angular momentum is particularly important for the thermal stability
of the coronae.  For instance, uniform rotation (unexpected in real
systems) would favour thermal instability, while (more realistic)
differential rotation with outward-increasing specific angular
momentum tends to stabilize.  A vertical gradient of $\Omega$
($\Omega$ decreasing for increasing $z$), might be expected if the
kinematic properties of the coronae reflect those of the neutral
extra-planar gas of disc galaxies \citep[e.g.][and references
therein]{Fra09}. In the absence of thermal conduction such a vertical
gradient, if sufficiently strong, might have a destabilizing effect
(compare baroclinic and barotropic models in
Fig.~\ref{fig:odeomegath}). However, such an instability is not
expected to occur in the presence of even highly suppressed
conductivity. In fact, the plots in the right-hand panel of
Fig.~\ref{fig:odeomegac} show clearly that conductive damping is
enhanced by the shear induced by differential rotation: even
non-axisymmetric perturbations that have initially long enough
wavelength---such that $\omegad(0)<0$---are effectively sheared and
therefore stabilized by thermal conduction, independently of whether
the distribution is barotropic (model 5b) or baroclinic (model 5d).

Altogether these considerations lead to the conclusion that, generally
speaking, differential rotation does not make galactic coronae
thermally unstable to non-axisymmetric perturbations. We cannot
exclude that thermal perturbations grow in specific cases, but a
definitive answer to the question of whether the corona of a given
galaxy is thermally unstable can be obtained only when a detailed
model of the corona, including its specific angular momentum
distribution, is available.

\section{Summary and conclusions}
\label{sec:con}

Motivated by the question of whether cool, pressure-supported clouds
can condense out of the hot coronae of disc galaxies, we studied the
problem of the thermal stability of rotating stratified fluids, in the
presence of radiative cooling and thermal conduction. As in the
non-rotating case, the time evolution of a perturbation depends on its
wavelength, so it is useful to distinguish short-wavelength
perturbations (such that the dissipative frequency $\omegad$ is
positive) from long-wavelength perturbations (such that the
dissipative frequency $\omegad$ is negative). We found that---against
either axisymmetric or non-axisymmetric perturbations---a {\it
  uniformly rotating}, convectively stable configuration is thermally
stable when $\omegad>0$ (damping by thermal conduction), but is
thermally unstable when $\omegad<0$. Similarly---against axisymmetric
perturbations---a {\it differentially rotating}, convectively stable
corona with outward-increasing specific angular momentum is thermally
stable when $\omegad>0$, but is thermally unstable when $\omegad<0$.
Non-axisymmetric perturbations in the presence of differential
rotation behave differently from the axisymmetric ones.  In the {\it
  absence of thermal conduction}, the combination of non-axisymmetry
of the perturbation and of differential rotation has a stabilizing
effect (turning instability into overstability), and barotropic
distributions tend to be more stable than baroclinic distributions.
In the {\it presence of thermal conduction}, stability replaces
overstability: in differentially rotating systems, the shear makes
conductive damping of non-axisymmetric disturbances particularly
effective for both barotropic and baroclinic distributions.

These results have been discussed in the context of the problem of the
growth of thermal perturbations in galactic coronae. The above
calculations allow to address unambiguously this question when applied
to a specific model of rotating galactic corona. Unfortunately, given
that the physical properties of the hot atmospheres of disc galaxies
(in particular, the distribution of specific entropy and angular
momentum) are very poorly constrained observationally, it is difficult
to be conclusive about whether these systems are prone to thermal
stability. Additional uncertainties come from magnetic fields, which
are expected to be present in real systems, but have been neglected
for simplicity in the studied models.  The question of the effect of
magnetic fields on the thermal instability is very interesting and
would require a full thermal-stability analysis of a rotating,
magnetized corona, for which our calculations may be a starting point.

Though limited by the mentioned uncertainties, the present work gives
further indications against the hypothesis that thermal instability is
important for galactic coronae.  In particular, though rotation can
destabilize against thermal disturbances very elongated in the
azimuthal direction, we argued that blob-like thermal perturbations
are unlikely to grow in a differentially rotating corona. This
finding, combined with previous results on non-rotating coronae (BNF),
suggests that the high-velocity clouds of the Milky Way did not form
spontaneously from small thermal perturbations in the Galactic corona,
but must be of external origin. A possibility is that at least the
seeds of these clouds are formed originally as a consequence of
stripping of gas-rich satellites or cosmic infall of cold gas.  These
cool gaseous seeds might then grow while travelling through the hot
Galactic corona \citep{Som06,Ker09}, for instance via turbulent mixing
\citep{Mar10}. Provided that the overdensities associated with these
accreted seeds are substantial (i.e. they are non-linear
perturbations), such a scenario for the formation of the high-velocity
clouds is not in contrast with the results of the present paper, in
which only linear perturbations are considered.

\section*{Acknowledgments}

I would like to thank Giuseppe Bertin and James Binney for helpful
discussions and useful comments on the draft.

\appendix

\section{Short-wavelength, low-frequency, axisymmetric perturbations are almost isobaric}
\label{app:isob}

Here we show that in a differentially rotating fluid short-wavelength,
low-frequency, axisymmetric perturbations are almost isobaric, in the
sense that $p/\pzero\ll \rho/\rhozero$, where $\pzero$ and $\rhozero$
are the unperturbed pressure and density, while $p$ and $\rho$ are the
pressure and density perturbations.  Let us consider the linearized
mass and momentum equations~(\ref{eq:axc}-\ref{eq:axm3}): eliminating
$\vR$, $\vz$ and $\vphi$, and we obtain
\begin{equation}
{p\over\pzero}={\alpha\over \beta}{\rho\over\rhozero},
\end{equation}
where 
\begin{equation}
\alpha\equiv {\omegahat^2\over \czero^2 k^2}-\i {\omegahat^2\over \omegabar^2}{\kR\ApR\over k^2}-2\i {\Omega\Omegaz\over \omegabar^2}{\kR\Apz\over k^2}-\i{\kz\Apz\over k^2} 
\end{equation}
and 
\begin{equation}      
\beta\equiv{\omegahat^2\over\omegabar^2}{\kR^2\over k^2}+2{\Omega\Omegaz\over\omegabar^2}{\kR\kz\over k^2}+{\kz^2\over k^2},
\end{equation}
with $\omegabar^2=\omegahat^2-2\Omega\OmegaR-2\Omega^2$.  In the limit
of short wavelengths and low frequencies adopted in
Section~\ref{sec:axisymm}, all terms of $\alpha$ are infinitesimal,
while $\beta$ is finite. It follows $p/\pzero\ll \rho/\rhozero$ and then
the perturbed equation of state $\rho/\rhozero=p/\pzero-T/\Tzero$ can
be approximated as $\rho/\rhozero\sim-T/\Tzero$ \citep[see][for the
  analogous calculation in the non-rotating case]{Tribble89}.

\section{Some properties of cubic equations}
\label{app:cubic}

We recall here a few properties of cubic equations. Let us consider
the equation
\begin{equation}
\ntilde^3+a \ntilde^2 + b \ntilde + c=0, 
\label{eq:cubic}
\end{equation}
in the unknown $\ntilde$, with coefficients  $a$, $b$ and $c$.
The associated discriminant is
\begin{equation}
\Delta=-4 a^3 c + a^2 b^2 - 4 b^3 +18 a b c -27 c^2.
\end{equation}
The roots of equation~(\ref{eq:cubic}) have the following properties:
\begin{enumerate}
\item[{\it - Number of real roots}.] If $\Delta \geq 0$ the equation has three real roots; if $\Delta < 0$
the equation has one real roots and a pair of complex conjugate roots.

\item[{\it - Routh-Hurwitz Theorem}.] All the roots have negative real parts if and only if the following conditions are satisfied:
\begin{equation}
a>0,\qquad
\left| \begin{array}{cc}
 a & 1\\
c & b \end{array} \right|>0,\qquad
\left| \begin{array}{ccc}
 a & 1 & 0\\
c & b & a \\
0 & 0 & c \end{array} \right|>0.
\end{equation}
\item[{\it - Vi\`ete's formulae}.] The three roots $\ntilde_1$, $\ntilde_2$ and $\ntilde_3$ satisfy Vi\`ete's formulae:
\begin{eqnarray}
\ntilde_1+\ntilde_2+\ntilde_3=-a,\\
\ntilde_1 \ntilde_2+\ntilde_2 \ntilde_3 +\ntilde_3 \ntilde_1=b,\\
\ntilde_1 \ntilde_2 \ntilde_3=-c.
\end{eqnarray}

\end{enumerate}


\end{document}